\documentclass[journal]{IEEEtran}
\usepackage{xcolor}

\usepackage{cite}
\ifCLASSINFOpdf
   \usepackage[pdftex]{graphicx}
\else
\fi
\usepackage{amsmath}
\usepackage{amssymb}
\usepackage{array}
\usepackage{url}
\begin{document}
%
% paper title
% Titles are generally capitalized except for words such as a, an, and, as,
% at, but, by, for, in, nor, of, on, or, the, to and up, which are usually
% not capitalized unless they are the first or last word of the title.
% Linebreaks \\ can be used within to get better formatting as desired.
% Do not put math or special symbols in the title.
\title{Towards Realistic Statistical Models of the Grid Frequency}

% author names and affiliations
% transmag papers use the long conference author name format.
\author{David Kraljic
\thanks{D. Kraljic is with Faculty of Electrical Engineering, University of Ljubljana, Ljubljana, Slovenia and with Comcom Trading d.o.o., Slovenia}
}

% \thanks{Manuscript received December 1, 2012; revised August 26, 2015. 
% Corresponding author: M. Shell (email: http://www.michaelshell.org/contact.html).}}

% The paper headers
% \markboth{Journal of \LaTeX\ Class Files,~Vol.~14, No.~8, August~2015}%
% {Shell \MakeLowercase{\textit{et al.}}: Bare Demo of IEEEtran.cls for IEEE Transactions on Magnetics Journals}
% The only time the second header will appear is for the odd numbered pages
% after the title page when using the twoside option.
% 
% *** Note that you probably will NOT want to include the author's ***
% *** name in the headers of peer review papers.                   ***
% You can use \ifCLASSOPTIONpeerreview for conditional compilation here if
% you desire.

% If you want to put a publisher's ID mark on the page you can do it like
% this:
%\IEEEpubid{0000--0000/00\$00.00~\copyright~2015 IEEE}
% Remember, if you use this you must call \IEEEpubidadjcol in the second
% column for its text to clear the IEEEpubid mark.

% use for special paper notices
%\IEEEspecialpapernotice{(Invited Paper)}

% for Transactions on Magnetics papers, we must declare the abstract and
% index terms PRIOR to the title within the \IEEEtitleabstractindextext
% IEEEtran command as these need to go into the title area created by
% \maketitle.
% As a general rule, do not put math, special symbols or citations
% in the abstract or keywords.
\IEEEtitleabstractindextext{%
\begin{abstract}
Increased share of renewable sources of energy in a power grid leads to larger deviations in grid frequency from the nominal value resulting in more challenging control and its modelling. In this paper we focus on the grid frequency for the power system of Great Britain because the large share of renewables makes it a template for other power grids in the future and because it exhibits peculiar statistical properties, such as long-term correlations in fluctuations, periodicity, bi-modality, and heavy tails in the distribution of the grid frequency. By modifications of the swing equation and the underlying noise statistics, which we justify qualitatively and quantitatively, we reproduce these peculiar statistical properties. We apply our model to realistic frequency response services and show our predictions outperform a standard swing equation model.
\end{abstract}

% Note that keywords are not normally used for peerreview papers.
}

% make the title area
\maketitle

% To allow for easy dual compilation without having to reenter the
% abstract/keywords data, the \IEEEtitleabstractindextext text will
% not be used in maketitle, but will appear (i.e., to be "transported")
% here as \IEEEdisplaynontitleabstractindextext when the compsoc 
% or transmag modes are not selected <OR> if conference mode is selected 
% - because all conference papers position the abstract like regular
% papers do.
\IEEEdisplaynontitleabstractindextext
% \IEEEdisplaynontitleabstractindextext has no effect when using
% compsoc or transmag under a non-conference mode.

% For peer review papers, you can put extra information on the cover
% page as needed:
% \ifCLASSOPTIONpeerreview
% \begin{center} \bfseries EDICS Category: 3-BBND \end{center}
% \fi
%
% For peerreview papers, this IEEEtran command inserts a page break and
% creates the second title. It will be ignored for other modes.
\IEEEpeerreviewmaketitle

\section{Introduction}
\label{sec:introduction}
The integration of renewable sources of energy into power grids is an ongoing trend which is mostly driven by an attempt to reduce the emissions of green-house gases during electricity production. The most prominent renewable sources are wind and solar power which are by their nature fluctuating and uncontrollable. High wind and solar share of power generation are typically associated with low effective inertia of the grid which leads to challenges in integrating them into a power system. Power grid with a significant share of renewables is susceptible to larger deviations in the grid frequency\cite{HOMAN2021116723, inertia_renewable} which leads to increase in procurement of reserve and ancillary services aimed at stabilising the grid. In recent years, due to the increased share of renewables, the typical size of deviations has been increasing \cite{HOMAN2021116723}, thereby increasing the probability of outages.

Realistic and detailed models of the grid frequency are therefore needed for future-proofing the grid. Modelling of the grid frequency underlies, for example, the estimation of the size of reserve and frequency response services that need to be procured as well as the estimation of the usage of assets, income, and risk of providers of such services.

Most models for the grid frequency start with the swing equation\cite{stevenson1994power}. {In order to include fluctuations inherent in the power system, the equation is upgraded to a stochastic differential equation resulting in a form of Ornstein-Uhlenbeck Process (OUP)\cite{stochastic_tpwrs}}.
The simplest models add a stochastic term directly to the swing equation and assume that fluctuations are adequately described by Gaussian noise \cite{g1, g2, g3}. {Another approach, following \cite{stochastic_tpwrs}, is to model the load fluctuations in the swing equation as an OUP (empirically supported by \cite{oup_load}), leading to two stochastic equations, one for the load and one for the grid frequency, resulting in a Brownian stochastic term for the swing equation \cite{double_peaked_1, double_peaked_2}.} As we will see in later sections, the properties of the `random' process underpinning the fluctuation in grid frequency exhibit a wealth of structure beyond Gaussian or Brownian.

Statistical modelling of grid frequency in literature usually focuses on the \emph{aggregated} values such as the probability density function (PDF). Some approaches study frequency as a \emph{timeseries}, which focuses on quantities such as the auto-correlation function (ACF). However, there has been little work on models that reproduce both the timeseries and aggregated properties.

The most salient feature of the grid frequency timeseries, the periodicity in the fluctuations, was identified to be caused by trading on the market \cite{trading_period}.
Reference \cite{data_driven} attempts to model these periodic effects, which show in the ACF, by the deterministic and periodic imbalance generated by the market while using Gaussian noise for the stochastic part. This reproduced the initial exponential drop-off of the ACF and the periodic peaks but had a mismatch in the tails and at the centre of the PDF for the grid frequency. 
Reference \cite{decentral} shows that the tails of the PDF for the grid frequency in continental Europe are heavier than Gaussian and finds that sourcing the noise from L\'{e}vy stable distributions improves the fit in the PDF tails, but now over-estimates the tails. Reference \cite{NATUREFREQ} improves on the goodness of fit of the PDF by solving the swing equation with Gaussian noise that has the mean and variance changing slowly with time, leading to q-Gaussian final distributions. Reference \cite{PhysRevResearch.2.013339} combines this noise model with the insight that the periodicity of the electricity market is responsible for the periodicity of the ACF leading to good quantitative fits to the PDF in the tails and rough qualitative agreement for the ACF.

{Grid frequency PDFs for different synchronous areas can roughly be separated in single-peaked and multi-peaked distributions \cite{grideye}. Reference \cite{double_peaked_2} has used the stochastic formulation of \cite{stochastic_tpwrs} and identified the governor deadband as the main ingredient for succesfully reproducing the double-peakedness of the frequency PDF for Great Britain. Similarly, detailed modelling of various frequency controllers identified the deadband as the primary reason for double-peakedness of the Irish grid \cite{double_peaked_1}. Reference \cite{effect_inertia} has further investigated the parameters in the swing equation (with an additional load OUP) to show that double-peakedness can be robustly removed with synthetic inertia.}

In contrast to statistical modelling of the grid frequency, one can attempt to predict it for some time frame in the future. Reference \cite{freq_predict} uses predictive models based on nearest-neighbour techniques, whereas \cite{freq_predict_ar} uses auto-regressive models. To avoid purely statistical analysis some properties of the grid frequency fluctuations can be explained by physical reasons. For example, \cite{wind_non_gauss} finds that wind profiles and particular locations of power injection exacerbate the non-Gaussianity of frequency fluctuations.

We select the power system of Great Britain (GB) for studying the statistical properties of grid frequency fluctuations. The GB system is its own synchronous area, so it can be studied in isolation. It is small (compared to e.g. the European Continental Synchronous Area), leading to low inertia, and has a large share of renewable energy\footnote{See e.g. \url{https://www.iea.org/data-and-statistics}}. Therefore, it can be considered as a template for future grids elsewhere. {Another reason for selecting the GB grid frequency is the fact that it exhibits non-trivial properties both in ACF \emph{and} PDF. Therefore, the methods in this paper that successfully model both can be extended to other synchronous areas with similar or simpler statistical properties. In the Appendix we show that similar statistical properties hold for several other synchronous grids.}

In Section \ref{sec:statist_summary} we present the statistical properties of the grid frequency we aim to reproduce with our modelling. In Section \ref{sec:phys_basis} we present the physical basis for the modelling, whereas in Section \ref{sec:pheno_model} we present a phenomenological model derived from the physical model. In Section \ref{sec:data} we describe the data used in the analyses. In Section \ref{sec:results} we show the results, discuss them, and use them in a real-world application.

\subsection{Our contribution}
\label{sec:contribution}
The overall contribution of the paper is a stochastic model of the grid frequency that reproduces the long-range dependence of fluctuations \emph{and} bi-modality of the distribution \emph{and} the heavy-tails of the distribution. All of these properties have not been captured so far in a single model and we achieve it with the same number of parameters as comparable models in literature. In detail:
\begin{enumerate}
    \item We show that the ACF has a power-law behaviour and we successfully model it with fractional noise. We show that the power-law cannot be reproduced by the usual non-fractional Gaussian or L\'{e}vy type noise within a swing equation framework.
    \item We show that the PDF has heavy tails and we successfully model it with noise that is both fractional and heavy-tailed (L\'{e}vy stable)
    {\item We empirically show that the effective damping factor (related to inertia and droop) is dependent on the grid frequency and describe a method of obtaining the values from the data. %We argue why this is expected for a realistic grid (besides governor deadbands). %Frequency dependent damping factor functional form due to e.g. governor deadbands is usually assumed (e.g. \cite{double_peaked_2, double_peaked_1}) and not extracted from the data.
    \item We model yearly energy and time requirements for two types of reserve services and show that our model outperforms the models based on Gaussian noise.}
\end{enumerate}

\section{The statistical properties of the GB grid frequency}
\label{sec:statist_summary}
\begin{figure}[ht!]
  \centering
  \includegraphics[width=\columnwidth]{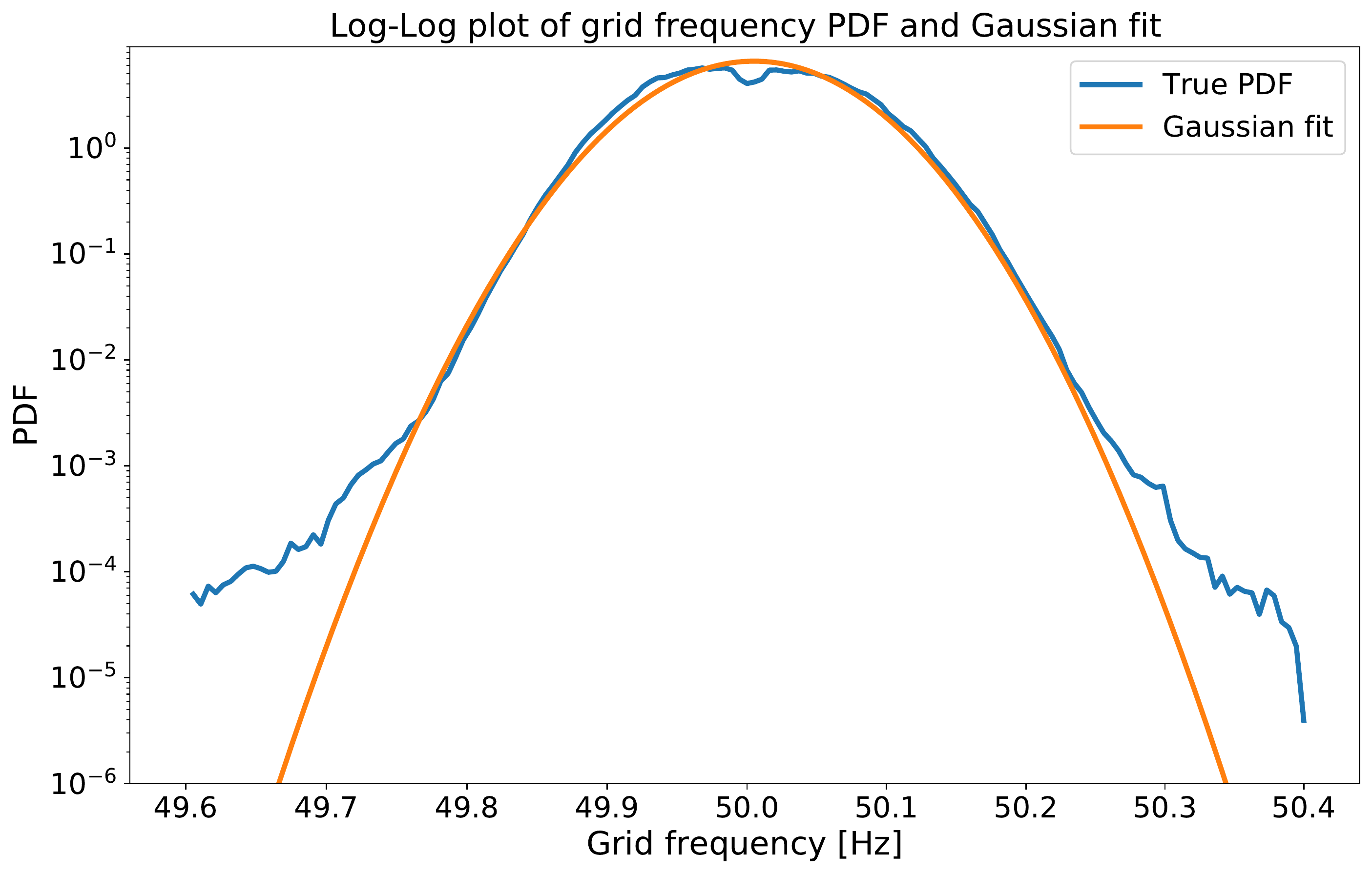}
  \caption{Log-Log plot of the PDF, with best-fit Gaussian overlaid. Note that the PDF above $\pm$0.2\,Hz exhibits heavier tails than Gaussian.}
  \label{fig:fat_tails}
\end{figure}

In this section we present the most salient statistical properties of the real GB grid frequency which we aim to reproduce in our models. The detailed discussion and modelling are deferred to later sections. The grid frequency fluctuates around its central value of 50\,Hz, so naively we would expect the most likely value to be 50\,Hz. However, the PDF of GB grid frequency (and many other grids) exhibits two peaks (Fig.\,\ref{fig:freq_pdf}), meaning that frequency spends more time slightly above or below the central value. The next property of the PDF reveals itself on its log-log plot (Fig.\,\ref{fig:fat_tails}) where we see that the tails of the distribution are heavier than what would be expected from Gaussian fluctuations. Usually, Gaussian fluctuations are expected as they arise via the Central Limit Theorem when many different fluctuating contributions sum up (e.g. in a power system). Heavy tails mean that the real grid frequency often exhibits large deviations. 

\begin{figure}[ht!]
  \centering
  \includegraphics[width=\columnwidth]{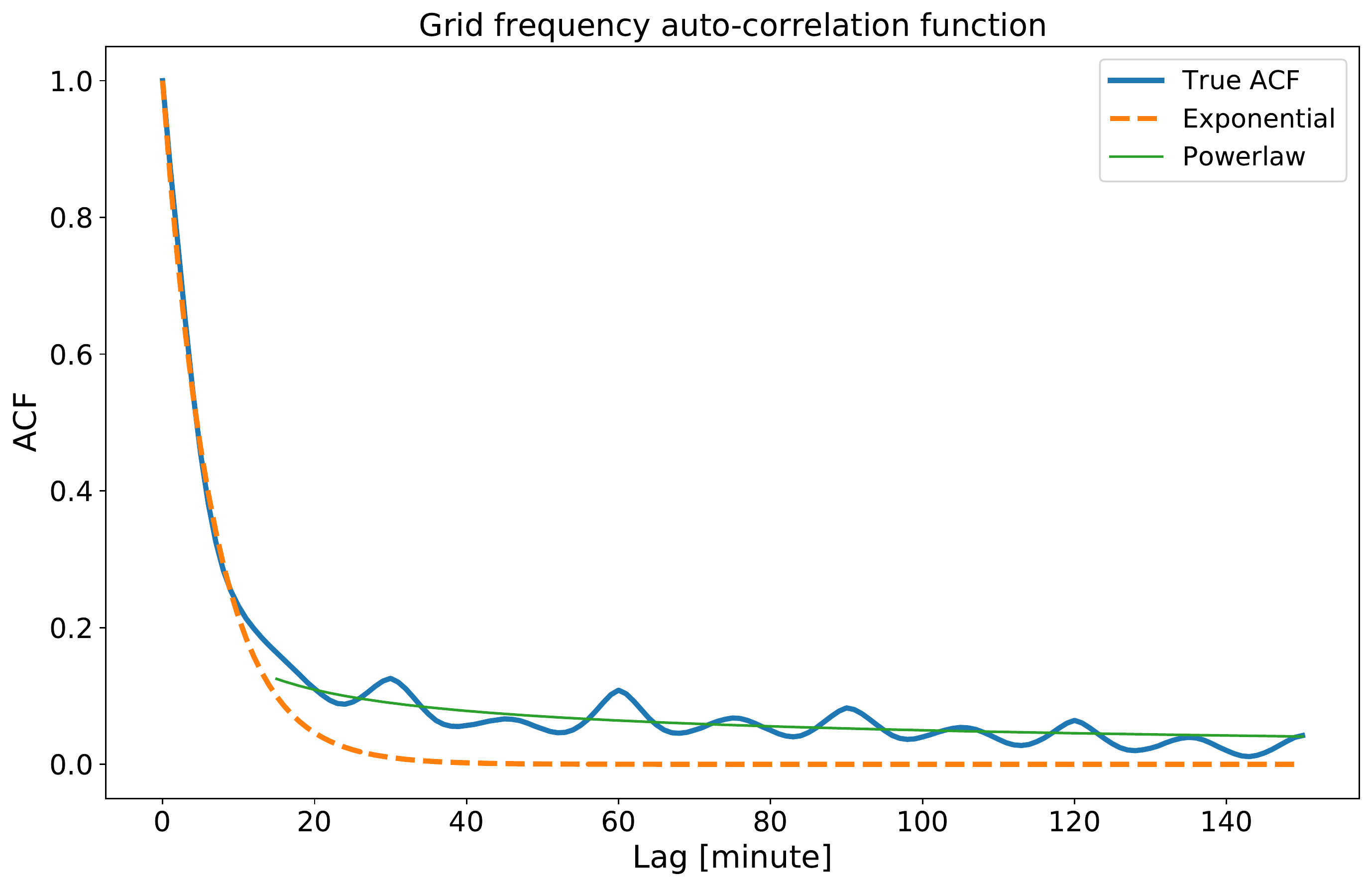}
  \caption{Autocorrelation of the GB grid frequency with lags up to 2.5 hours (150 minutes). Note the quick exponential decay initially followed by a slow power-law-like decrease later.}
  \label{fig:autocor_long}
\end{figure}

\begin{figure}[ht!]
  \centering
  \includegraphics[width=\columnwidth]{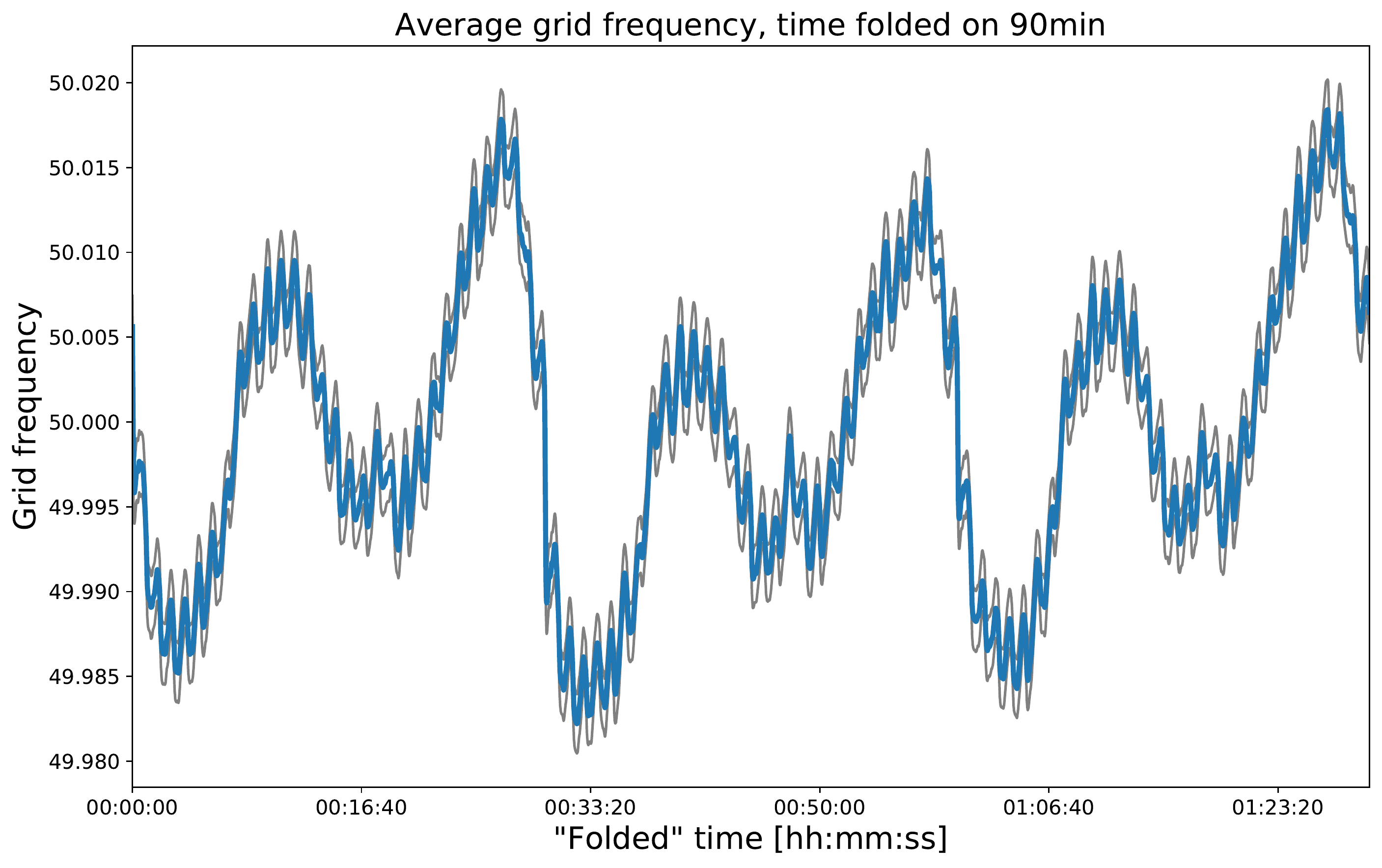}
  \caption{Phase dispersion periodogram -- average grid frequency with time `folded' on the period of 90 minutes. This means that, for example, the value for the time label 00:00:12 is the average of grid frequency at times 00:00:12, 01:30:12, 03:00:12, etc.. The two standard deviation bands are plotted in grey.}
  \label{fig:freq_fold}
\end{figure}

Deviation of the grid frequency away from the central value is not independent from one time instant to another, because the fluctuations driving the grid frequency, i.e. the noise, are adding up over time. The dependence over time is captured by the ACF (Fig.\,\ref{fig:autocor_long}). We would like to capture both the short term drop-off, that measures how strongly the value of grid frequency is pushed towards the center, and the long-term power-like tail, that signals that deviations away from the central value can persist for hours. We would also like to mimic the periodic nature in the ACF (Fig.\,\ref{fig:freq_fold}). Summarized properties that our model aims to capture are:

\begin{figure}[ht!]
  \centering
  \includegraphics[width=\columnwidth]{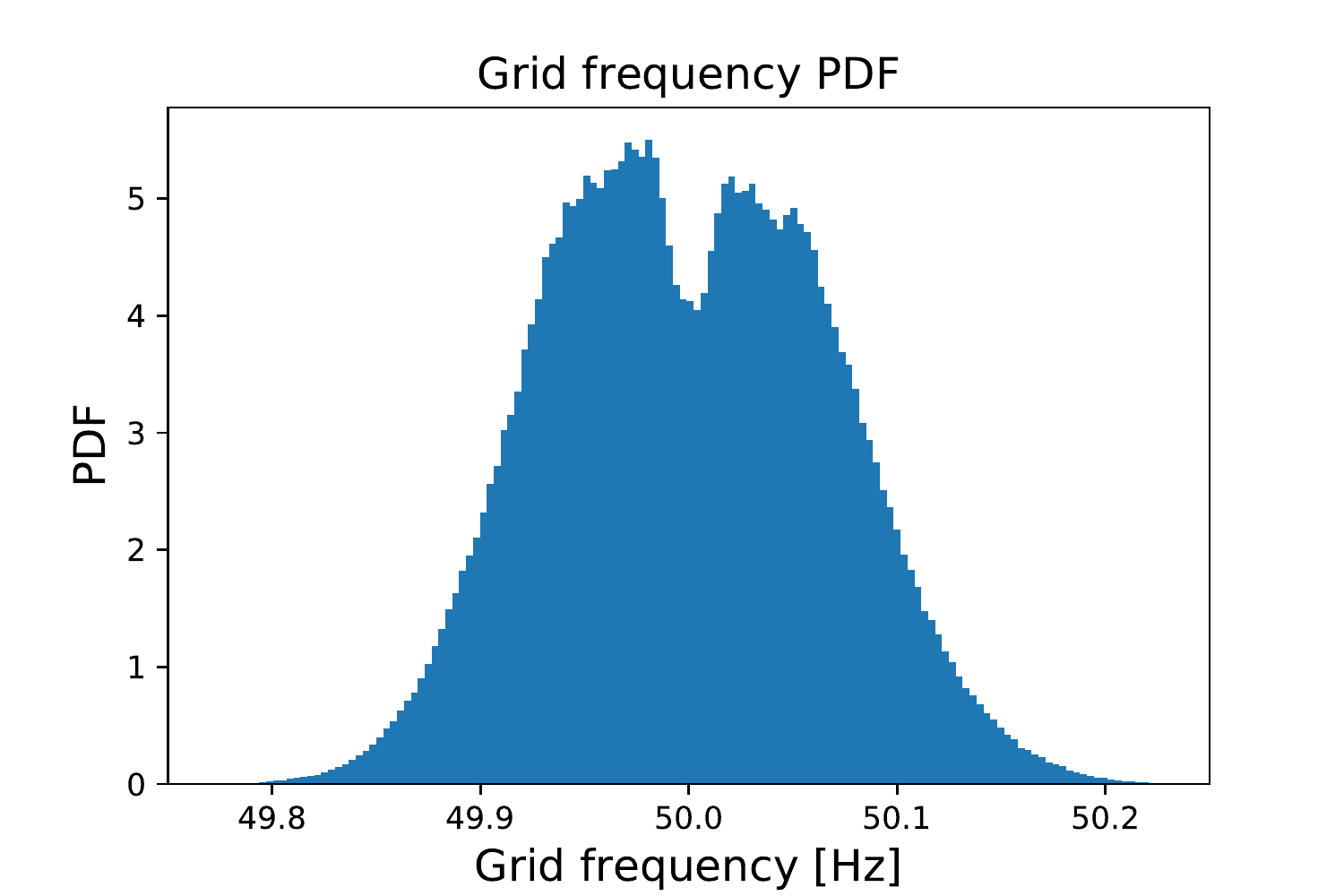}
  \caption{Histogram (PDF) of the GB grid frequency. Note the bi-modal (double-peaked) structure.}
  \label{fig:freq_pdf}
\end{figure}

\begin{figure}[ht!]
  \centering
  \includegraphics[width=\columnwidth]{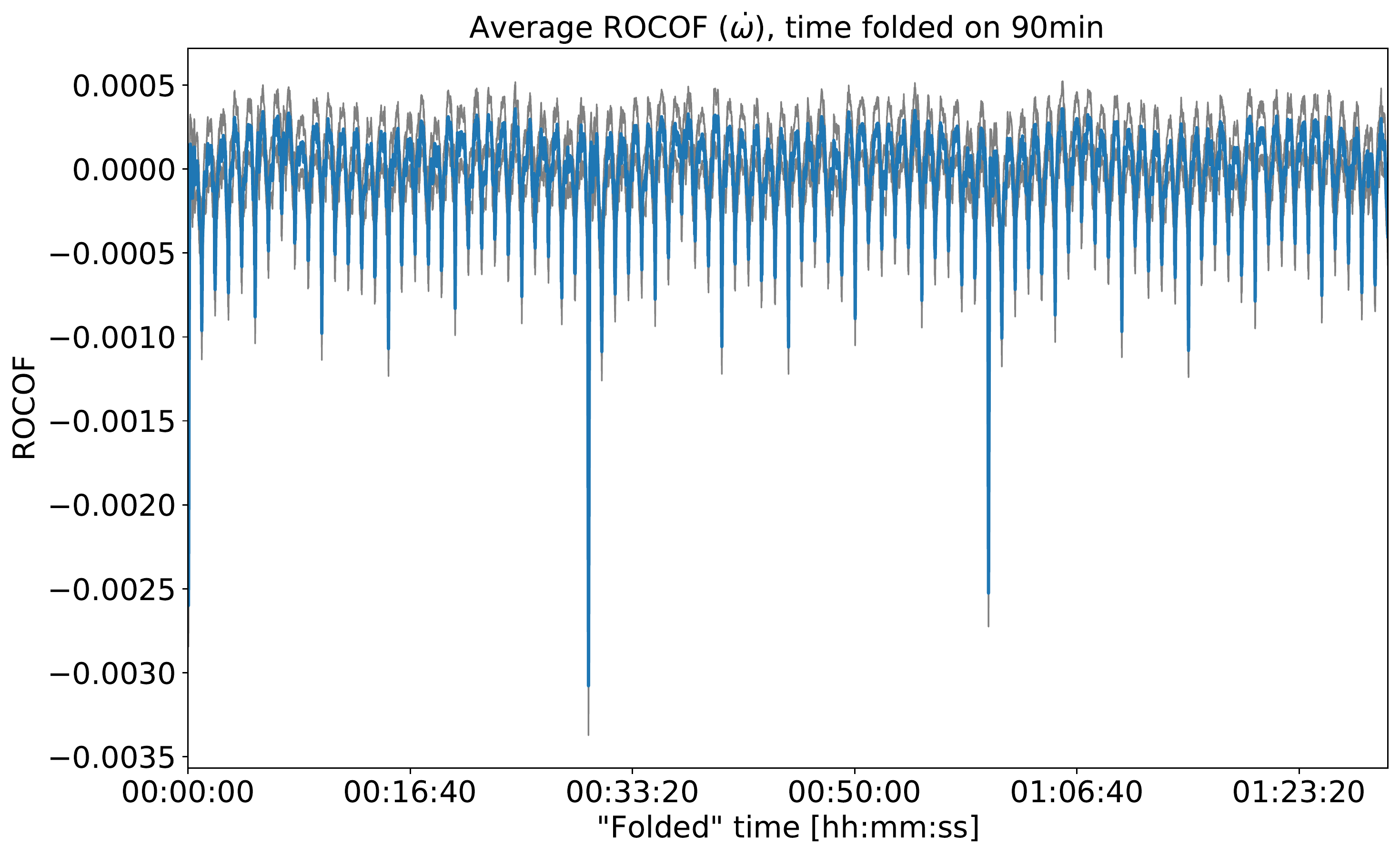}
  \caption{Phase dispersion periodogram -- average ROCOF (rate of change of frequency), time `folded' on the period of 90 minutes. The two standard deviation bands are plotted in grey.}
  \label{fig:rocof_fold}
\end{figure}

\begin{enumerate}
    \item The grid frequency PDF
    \begin{enumerate}
        \item The bi-modal (double-peaked) structure, Fig.\,\ref{fig:freq_pdf}
        \item The overall scale of fluctuations
        \item The heavy tails, Fig,\,\ref{fig:fat_tails}
    \end{enumerate}
    \item The autocorrelation function / phase dispersion periodogram
    \begin{enumerate}
        \item The exponential drop for short lags ($\lesssim$10min), Fig.\,\ref{fig:autocor_long}
        \item The periodicity of signal, Figs.\,\ref{fig:autocor_long}, \ref{fig:freq_fold}
        \item The second-by-second structure (within timescale of exponential drop), Figs.\,\ref{fig:freq_fold}, \ref{fig:rocof_fold}
        \item The power-law like long range correlation, Fig.\,\ref{fig:autocor_long}
    \end{enumerate}
\end{enumerate}

\section{The physical basis for the model}
\label{sec:phys_basis}
The electric grid consists of many interconnected generators and consumers of power, each of which can be considered as an oscillator producing/consuming power at the frequency close to the nominal grid frequency $\Omega$ (e.g. 50\,Hz). The following analysis closely follows \cite{GRIDKURAMOTO}.
The phase angle at a generator $i$ is:
\begin{equation}
    \theta_i = \Omega t + \tilde{\theta}_i
\end{equation}
where $\tilde{\theta}_i$ is the small phase perturbation relative to the whole grid.
The power of the generator is then:
\begin{equation}
    P_{i} = P^{\text{dissipated}}_i + P^{\text{accumulated}}_i + P^{\text{transmitted}}_i
    \label{eq:power_sum}
\end{equation}
where the dissipated power is due to friction, the accumulated power is stored in rotational kinetic energy, and the transmitted power is due to the phase difference between this generator and the rest of them:
\begin{align}
    P^{\text{dissipated}}_i \propto \left(\dot{\theta}_i\right)^2, \\
    P^{\text{accumulated}}_i \propto M_i \frac{d}{dt} \left(\dot{\theta}_i\right)^2, \\ P^{\text{transmitted}}_i \propto \sum_{j \neq i} K_{i,j} \sin\left(\theta_j - \theta_i\right),
\end{align}
where $M_i$ is the generator moment of inertia, $K_{i,j}$ contains the information on the topology  and transmission capacity of the grid (which generators are connected and the amplitude of the power flow between them), and is symmetric in indices.

{Expanding each term on the right-hand-side of (\ref{eq:power_sum}) to first order in perturbations $\tilde{\theta}_i$ and dividing everything by the moment of inertia $M_i$ results in:
\begin{align}
    \ddot{\widetilde{\theta}}_i = p_i - \alpha_i \dot{\widetilde{\theta}}_i + \sum_{j \neq i} k_{i,j} \sin\left(\theta_j - \theta_i\right)
\end{align}
where  parameter $p_i=P_i/M_i$ corresponds to the power of generator $i$, $\alpha_i$ to the magnitude of friction, and $k_{i,j}=K_{i,j}/M_i$ to the amount of power that can be transmitted between generator $i$ and $j$.}

The overall grid frequency $\omega$ is the nominal value $\Omega$ plus the sum (weighted by the inertia of each generator) of all individual perturbations:
\begin{align}
    \omega = \Omega + \sum_i M_i \dot{\widetilde{\theta}}_i / \sum_{i} M_i
\end{align}
which results in the `swing equation':
\begin{align}
    \dot{\omega} = - \gamma \omega + \pi\left(t\right)
    \label{eq:freq_phys}
\end{align}
where the first restoring term on the right-hand-side contains is pulling the grid frequency towards the nominal value $\Omega$ (which we dropped for convenience and without loss of generality) and the second term $\pi\left(t\right)$, which perturbs the grid frequency, contains the sum of all generated and consumed powers at each generator:
\begin{align}
    \pi\left(t\right) \propto \sum_{i} M_i P_i
\end{align}
The swing equation (\ref{eq:freq_phys}) is of the form as in e.g. \cite{double_peaked_1, NATUREFREQ}. There is an equivalent formulation, as in e.g.\cite{double_peaked_2, stochastic_tpwrs}:
\begin{align}
    2H\dot{\omega} = - \alpha \omega + \pi\left(t\right),
    \label{eq:inert_droop}
\end{align}
where $H$ is the aggregated inertia constant and $\alpha$ is the inverse aggregated droop coefficient. Note that if we increase or decrease the inertia $H$, droop $\alpha$, and the size of the noise term $\pi\left(t\right)$ by the same factor, the dynamics of grid frequency as described by the swing equation does not change. Therefore, when simulating the swing equation, there is a degeneracy among these parameters. However, we can constrain the ratios, e.g. $\gamma = \alpha / 2H$, so we choose the swing equation parametrization as in (\ref{eq:freq_phys}).\footnote{Inertia constant and droop can be estimated if the parameter degeneracy is broken by a known size of the power fluctuation $\pi\left(t\right)$, e.g. during the first few seconds of a sudden outage of a large power plant.}

\section{The phenomenological model}
\label{sec:pheno_model}
The physical model of (\ref{eq:freq_phys}) forms the basis of our modelling of the grid frequency. The physical model contains two unknowns -- the perturbation term $\pi\left(t\right)$ and the decay factor $\gamma$. In this section we consider what determines the values and structure of these unknowns and how they can be empirically modelled.

\subsection{The nature of perturbation term $\pi\left(t\right)$}
\label{subsec:pi}
In \cite{NATUREFREQ} it is assumed that the powers sum to zero and all that remains is unstructured noise (either of Gaussian or L\'{e}vy type) of unknown origin or cause. Specifically, it is assumed that the noise at two different times is uncorrelated. In reality, the powers do not sum to zero at each time instant, as the residual is what drives grid frequency away from the nominal value, and the term $\pi\left(t\right)$ contains a wealth of stochastic and deterministic structure. Of course, the residual of the sum of powers of all rotors can be in first instance approximated as a noise term.

\paragraph{Random part} 
The random part of $\pi\left(t\right)$ is caused, for example, by the variability in production of renewable generation, the forecast error in demand, the forecast error of the market regarding imbalance, as well as the effect of faults in the grid (generator tripping). The information on the forecast errors is not available immediately, the error is not rectified immediately (e.g. weather forecast run only a few times daily), and the information does not spread efficiently to all the active participants in the grid (generators, electricity system operators (ESO), traders). This means that the random part of the frequency deviation might exhibit long term dependence.

\paragraph{Non-random part}
The non-random part of $\pi\left(t\right)$ is influenced, for example, by the scheduled decisions of generators, the balancing actions of the ESO, the developments on the power market, the automatic activation of reserve services, generators speculating on the imbalance price, and generators misreporting their physical position. All of the listed things are tied to the common structure -- the settlement of imbalances, which is performed at half-hourly intervals in GB, and the market contracts that are structured into half-hourly, 1-hour, 2-hour, and 4-hour blocks. Therefore, we expect deterministic structure to appear and be periodic at these time scales.

The non-random part of $\pi\left(t\right)$ also contains responses to the system imbalances by active participants in the grid, which can all be described as terms of the restoring form like $-\gamma^{\text{parti.}}\omega$. For example, the ESO is actively pursuing the balance of the electric system, so its actions can be described by a restoring term. The market, through matching the supply and demand, is also driving the system towards balance and can be described by a restoring term. The reserve (a.k.a frequency response) triggered at certain deviations from the nominal frequency and proportional to the size of deviation can be faithfully modelled by a restoring term. Imbalance speculating generators that intentionally produce imbalance to collect payment according to the imbalance price also drive the system towards stability, and consequently frequency towards the nominal value.

Therefore, after removing the terms of the restoring form, the non-random part of the noise term in (\ref{eq:freq_phys}) can be decomposed into the pure noise term $\xi(t)$ (of unit variance) with the mean $\mu(t)$ and variance $\sigma(t)$ being periodic functions determined by e.g. the periodicity of the market (as in Fig.\,\ref{fig:freq_fold}):
\begin{align}
    \pi(t)= \mu(t) + \sigma(t) \cdot \xi(t)
    \label{eq:noise_decompose}
\end{align}

\subsection{The nature of damping factor $\gamma$}
\label{subsec:gamma}
{The restoring term $- \gamma (\omega - \Omega)$ in (\ref{eq:freq_phys}) is in part physical in origin and is determined by the moment of inertia  and droop of all the generators in the grid (\ref{eq:inert_droop}).}

In most studies so far the damping factor $\gamma$ has been assumed to be constant in time, with the exception of \cite{NATUREFREQ}. In fact, due to the varying share of renewable generation in the grid, the moment of inertia of the grid is lower during times of high share of renewable generation. For a review, see \cite{wind_inertia, ercot_inertia}. Therefore, we expect $\gamma$ to be a slowly (compared to $\pi\left(t\right)$) varying function of time. 

The restoring part of (\ref{eq:freq_phys}) has multiple contributions in addition to the physical in origin. As discussed in Sec.\,\ref{subsec:pi}, the active participants in the grid (generators, ESO, traders, reserve services) also contribute terms of the restoring form. Therefore, the restoring term is described by the sum of all contributions as an overall effective damping factor $\gamma^{eff}$. 

{We expect the effective damping factor $\gamma^{eff}$ to have a dependence on the value of grid frequency. The main physical reason are governor deadbands. Other restoring terms non-physical in origin also contain `deadbands'. For example, ESO does not react to the smallest system imbalance (as manifested in frequency deviation) and instructs power plants only when deviation is big enough. Additionally, reserve services are only triggered above/below a certain threshold. All of these effects combine into strong $\omega$ dependence of the effective damping term $\gamma^{eff}$:}
\begin{align}
    \dot{\omega} = - \gamma^{eff}\left(t, \omega\right) \cdot \omega + \pi\left(t\right)
    \label{eq:freq_full}
\end{align}
where $\pi\left(t\right)$ now contains only the random and periodic components (\ref{eq:noise_decompose}). In this paper we will not focus on the time dependence of $\gamma^{eff}$ but instead on the dependence on $\omega$.

\subsection{Modelling the noise $\xi\left(t\right)$}
The restoring term of (\ref{eq:freq_phys}) is responsible for the initial exponential decay of the ACF. In the case where $\gamma^{eff}$ is constant and $\pi\left(t\right)$ is uncorrelated Gaussian noise, the model is known as  the Ornstein–Uhlenbeck\cite{ou_process} (or Vasicek\cite{VASICEK1977177}) model and has an autocorrelation function that decays exponentially with the decay factor $\gamma^{eff}$. 

Extending such a model by heavy-tailed noise sourced from e.g. a L\'{e}vy stable distribution does not change the exponential drop of the ACF. This can be seen by integrating (\ref{eq:freq_phys}) and calculating the correlator $\langle \omega(t) \omega(t')\rangle$\cite{psd}. We use the fact that for Gaussian and L\'{e}vy type noise $\langle \xi(t) \xi(t') \rangle \propto \delta(t-t')$, where $\delta(t-t')$ is the Dirac delta function. That is, the noise two-point correlation function is non-vanishing only when evaluated at the same instant\cite{levy}. Therefore, the long-range dependence seen empirically in the ACF (see Fig.\,\ref{fig:autocor_long}) cannot be produced by these types of noise.

\paragraph{Long-range dependence via Fractional noise}
Long term dependence, power-law behaviour, and long-term memory processes are frequently described using `fractional noise processes' \cite{mandelbrot,beran1994statistics,hurst_psd}. For example, Fractional Gaussian Noise is a generalisation of Gaussian noise for which the correlation of the noise term at two different time instants does not vanish and thus leads to long-range dependence in the ACF.
\begin{table}[!t]
\caption{Types of noise and their properties\label{tab:table1}}
\centering
\begin{tabular}{|c||c|c|}
\hline
Property & Gaussian / L\'{e}vy & Fract. Gaussian / L\'{e}vy \\
\hline
\hline
Mean $\langle\xi(t)\rangle$ & 0 & 0 \\
\hline
Correlation $\langle\xi(t)\xi(t')\rangle$ & $\propto \delta(t-t')$ & $\propto |t-t'|^{2H-2}$ \\
\hline
Tails & normal / fat & normal / fat \\
\hline
\end{tabular}
\end{table}

Long-range behaviour of fractional noises is characterised by a single parameter $H$, called the Hurst exponent. Uncorrelated behaviour is reproduced for $H=1/2$,  $1> H >1/2$ results in positive long-range autocorrelations, and $0<H<1/2$ results in short-term autocorrelations. 

Heavy tails can also be modelled within the fractional noise framework, where the underlying noise is sourced from a L\'{e}vy stable distribution \cite{nolan2020univariate}, thus achieving both the long-range correlations and large point-wise fluctuations. This type of noise is termed Fractional L\'{e}vy Noise (FLN) \cite{frac_levy}. We are interested in noise distributions without skew. Therefore the L\'{e}vy stable distributions we consider are characterized by one parameter $\alpha \in (0,2]$ which controls the fatness of tails, where values below 2 correspond to heavy tails. 

Standard uncorrelated Gaussian noise leading to OUP swing equation, used in e.g. \cite{double_peaked_1, double_peaked_2}, is reproduced for parameter combination $H=1/2$ and $\alpha=2$.
  
Fractional noise processes have found application in a variety of fields (e.g. economics \cite{fbm}, hydrology \cite{hydro}, engineering \cite{engineer}...).
The properties of different noise types are summarized in Table \ref{tab:table1}.

\section{The data}
\label{sec:data}
The data on GB grid frequency was obtained from the electricity system operator (ESO) in Great Britain\cite{ng_data} and it consists measurements of the grid frequency floored to three decimal places with granularity of 1 second. We study the data from 2014 to 2020 inclusive.

%Note that the flooring creates `quantization' error of approximate variance $\Delta^2/12$ where $\Delta$ is the rounding step size. This error is particularly significant when $\dot{\omega}$ is estimated from the data by differencing the subsequent values, which leads to doubling of the quantization variance in the worst case. Note that this variance is comparable to the typical values of $\dot{\omega}$.  Therefore, using $\dot{\omega}$ in analyses (e.g. using regression on Eq.\,\ref{eq:freq_full}) to estimate the damping factor) should be avoided.

\section{Results and Discussion}
\label{sec:results}
\subsection{Autocorrelation function: the exponential part}
The autocorrelation function has an initial exponential decay and a long-range power-law like part (see Fig.\,\ref{fig:autocor_long}). For lags $\lesssim$ 15\,min the ACF can then be modelled as $\text{ACF}\left(t\right) \sim \exp\left(-\gamma^{eff} t\right)$. The parameter $\gamma^{eff}$ can be extracted by straight line fits on a semi-log plot of the ACF. The results of the fit is $\gamma^{eff} = (5.5 \pm 0.1 \text{\,min})^{-1}$.

\subsubsection{Dependence of $\gamma^{eff}$ on $\omega$}
As explained in Sec.\ref{subsec:gamma} the \textit{effective} damping factor is expected to have a dependence on the value of the grid frequency, i.e. $\gamma^{eff}(\omega)$.

%A straightforward approach to estimating $\gamma^{eff}(\omega)$, that is $\gamma^{eff}$ tabulated by frequency bins, would be to use (\ref{eq:freq_full}), select data within the target frequency bin and regress $\dot{\omega}$ on $\omega$ and extract $\gamma^{eff}$ as the coefficient. However, this would lead to biased estimates for the mean and variance of $\gamma^{eff}$. The significant and long-term autocorrelation in the residuals (the error term of (\ref{eq:freq_full})) leads to biased estimates, which can in principle be corrected for by detailed modelling of the covariance matrix for the error term \cite{linreg_correction}). The rounding or flooring creates a quantization error of approximate variance $\Delta^2/12$ where $\Delta$ is the rounding step size,  which can lead to additional biases 

%Therefore, the parameter $\gamma^{eff}$ is estimated from the autocorrelation function of $\omega$, where it corresponds to the slope in the log plot of the ACF.
%We calculate the ACF within a frequency bin by restricting one of the two factors in the ACF to be within the target frequency bin. That is, we study quantity of the form $\langle \omega(t| \omega\in \textsc{bin})\omega(t')\rangle$. 

\begin{figure}[ht!]
  \centering
  \includegraphics[width=\columnwidth]{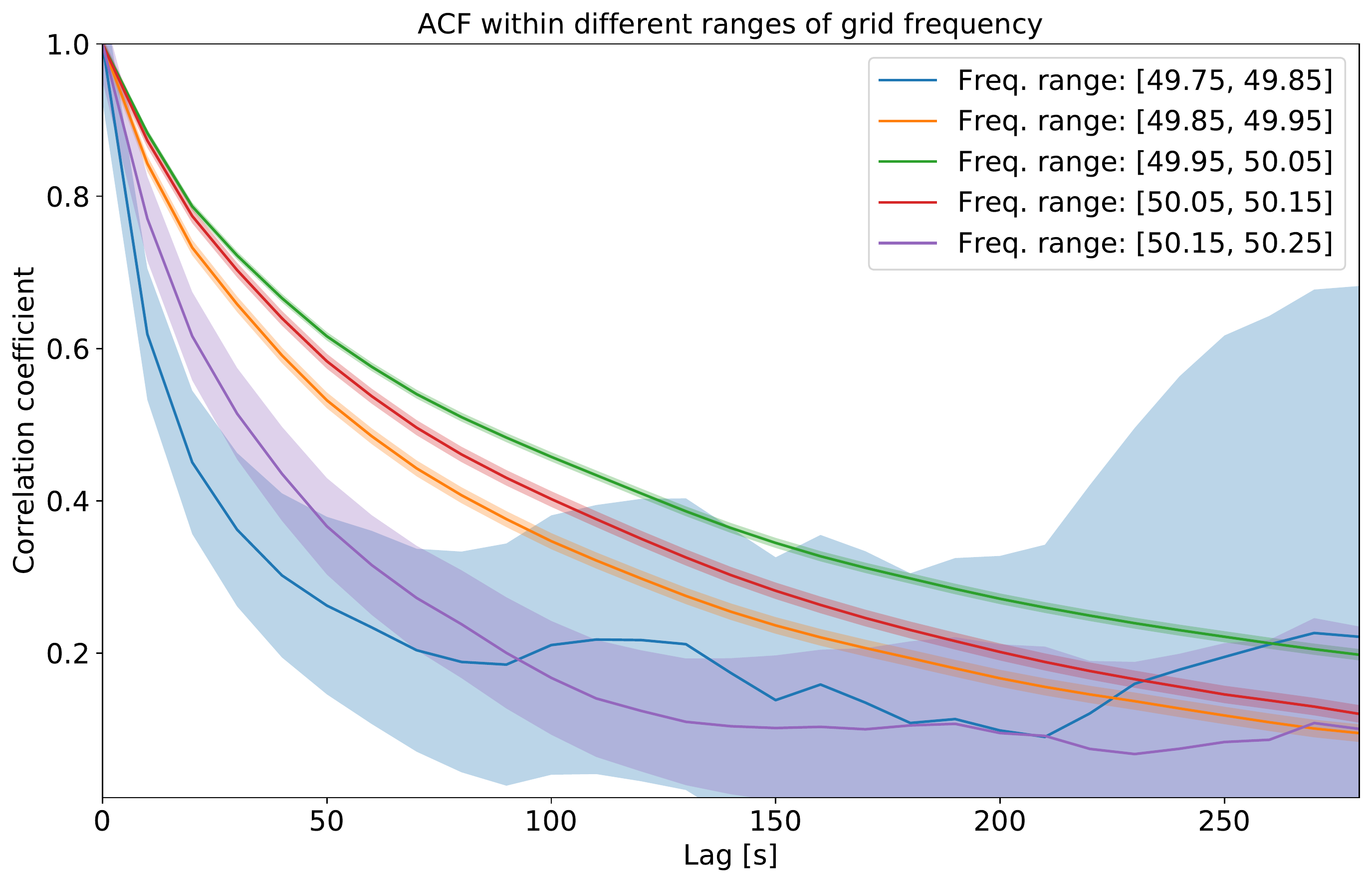}
  \caption{{ACF dependence on the grid frequency, in the regime where exponential drop-off is a good description. Note that when frequency is closer to 50\,Hz we have stronger auto-correlations caused by lower $\gamma^{eff}$. Note that the error in ACF increases with lag distance (as there are fewer independent data points to estimate the ACF) and increases for bins far away from the nominal value, as the number of data points in those bins is small.}}
  \label{fig:ACF_vs_freq}
\end{figure}

{To show this effect is present in the data we split the datasets in frequency bins (e.g. in bins of 0.1\,Hz). Then we calculate the ACF for data falling in each of the bins and plot the results in Fig.\,\ref{fig:ACF_vs_freq}. % and in Fig.\,\ref{fig:gamma_vs_freq}. 
We plot the first few lag minutes in the ACF since the exponential drop-off there is controlled by $\gamma^{eff}$. We see that for frequencies closer to 50\,Hz the fluctuations are correlated stronger and for longer time. This means that $\gamma^{eff}$ is smaller closer to 50\,Hz as was expected (Sec.\ref{subsec:gamma}).
The values for $\gamma^{eff}(w)$ can finally be extracted as slopes to semi-log plot of Fig.\,\ref{fig:ACF_vs_freq}.}
 %Here, special care must be taken as the error in the ACF estimate does not scale as $1/\sqrt{N}$ where $N$ is the number of data points, due to significant autocorrelation which reduces the effective number of data points. We employ a simple correction \cite{corr_correction} to obtain the effective number of data points $N_{eff}$ for each estimate of the ACF.

% \begin{figure}[h!]
%   \centering
%   \includegraphics[width=\columnwidth]{gamma_vs_w_final.pdf}
%   \caption{$\gamma_{eff}$ dependence on the grid frequency bin}
%   \label{fig:gamma_vs_freq}
% \end{figure}

\subsection{Autocorrelation function: the power-law part}
For the autocorrelations above roughly 15\,min the behaviour of the ACF can be described as a power law, $\text{ACF}\left(t\right) \sim t^{-\alpha}$. The parameter $\alpha$ can be extracted by straight-line fits on the log-log plot of the ACF. The result of the fit is $\alpha=-0.5\pm0.1$.

\begin{figure}[ht!]
  \centering
  \includegraphics[width=\columnwidth]{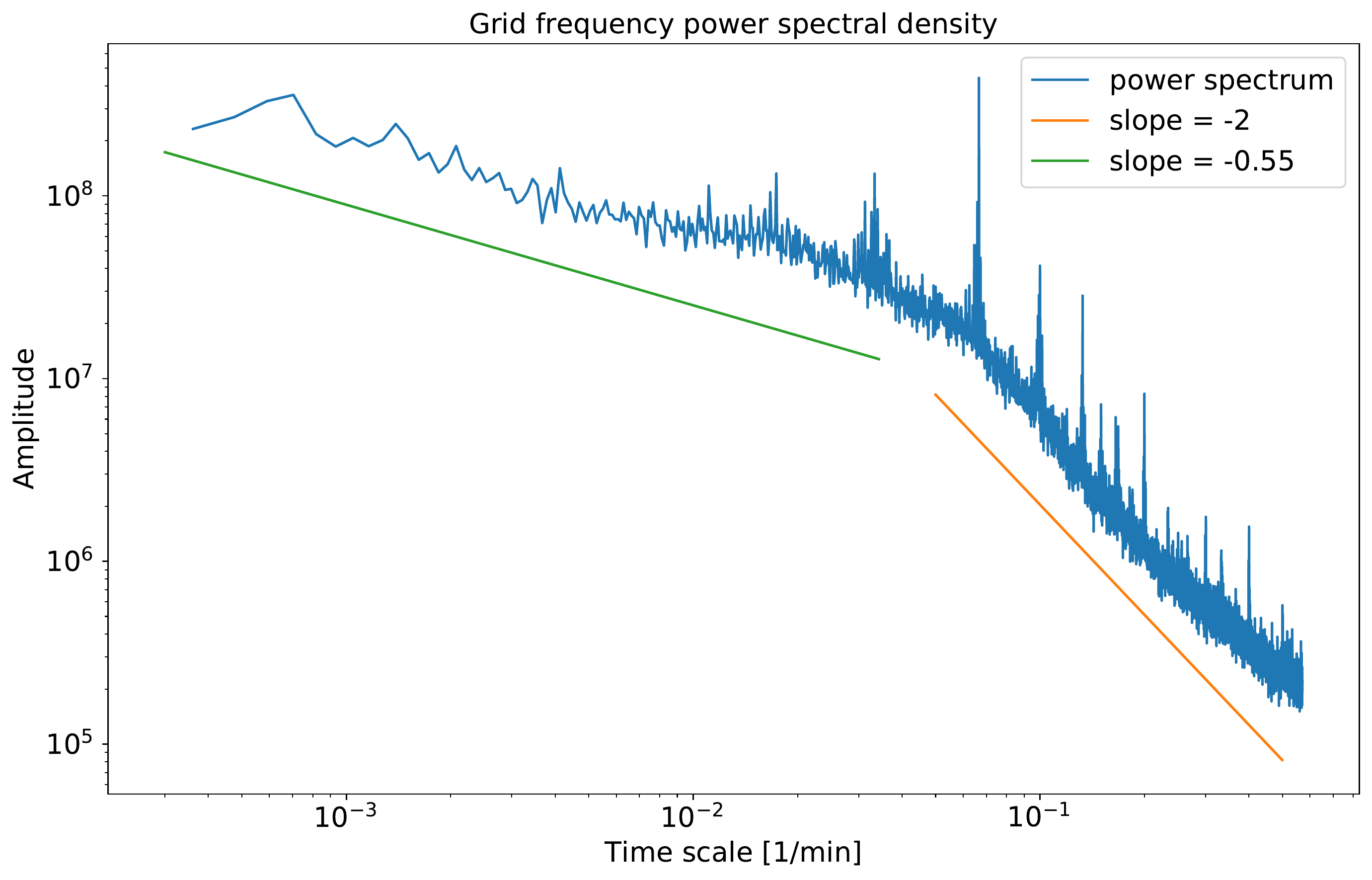}
  \caption{Power spectral density of the grid frequency stochastic process. Note the Brownian/Gaussian scaling (slope=-2) at short times and fractional Brownian scaling at long times. For purely Gaussian noise the spectral density slope should be 0 at long times. Note the peaks due to periodicity.}
  \label{fig:psd}
\end{figure}

We model long-range, power-law correlations, using fractional noise. The stochastic noise is described by one parameter, the Hurst exponent $H$, which leads to power-law ACF of $t^{2H-2}$ \cite{hurst_power_law, hurst_v_good}. Grid frequency power-law fit leads to Hurst exponent of 0.75 ($2H-2=-0.5$), suggesting strong long-term dependence. An alternative approach to demonstrate short time exponential decay in the ACF and long time power law is by studying the power spectral density of the process, which we plot in  Fig.\,\ref{fig:psd}. Power spectral density for the swing (OU, Vasicek) type of equation as (\ref{eq:freq_phys}) is described by the following expression \cite{psd}:

\begin{align}
    PSD(k)\sim \frac{\langle \tilde{\pi} \tilde{\pi}^{*}\rangle}{\gamma^2 + k^2}
    \label{eq:psd}
\end{align}
where $k$ is the inverse of the time scale, $\gamma$ is the decay constant, and $\langle \tilde{\pi} \tilde{\pi}^{*}\rangle$ is the noise power spectrum. For Gaussian/L\'{e}vy noise this is just a constant (Fourier transform of the delta function). However, for fractional noise this term scales as $\langle \tilde{\pi} \tilde{\pi}^{*}\rangle \sim k^{-2H + 1}$ \cite{hurst_psd, hurst_v_good}. From the slope ($-0.55 \pm 0.04$) at long times (small $k$) we show, that the underlying noise process is not Gaussian (slope should be 0) and obtain the Hurst exponent of about 0.75, consistent with estimation from the ACF.

\subsection{Probability density function: heavy tails}
We generate heavy-tailed long-range Fractional L\'{e}vy Noise using our own efficient implementation\footnote{\url{https://github.com/davidkraljic/fractional_levy_motion}} of \cite{matlab_simulation}. Parameter $\alpha$ that controls the fatness of tails can be in principle estimated from the data by e.g. a Hill estimator\cite{hill_estimator}. However, the grid frequency PDF is close to Gaussian (deviations away from Gaussian only above $\pm$0.2\,Hz) suggesting that $\alpha$ is close to 2 (Gaussian case). In this parameter region estimators are unreliable\cite{nolan2020univariate}. Therefore, we perform a direct search via parameter sweep to find $\alpha=1.975$ that best matches the tails of the GB grid frequency PDF (see Fig.\ref{fig:hist_simulation_tails}).

\subsection{Simulation results}

Equations (\ref{eq:freq_phys}) and (\ref{eq:freq_full}) are stochastic differential equations, corresponding to a mean-reverting process. Numerical solutions to (\ref{eq:freq_phys}) and (\ref{eq:freq_full}) are obtained using the Euler–Maruyama method\cite{simulating}.
The equations are solved using \textsc{numpy} software package\cite{numpy}.

We compare three models: a simple model based on the swing equation (\ref{eq:freq_phys}) with non-fractional Gaussian noise covers the majority of previous modelling attempts (see Section \ref{sec:introduction}), a detailed existing model for the GB grid frequency \cite{double_peaked_2}, and our model based on (\ref{eq:freq_full}) with fractional L\'{e}vy noise for $\xi(t) $ and $\omega$ dependent $\gamma^{eff}$. The periodic component of the noise term $\mu(t)$ and scale of fluctuations $\sigma$ are both taken from actual data which is summarised in Fig.\,\ref{fig:rocof_fold}. We model $\gamma^{eff}$ with three parameters (as in e.g. \cite{double_peaked_1, double_peaked_2}): 
\begin{align}
    \gamma^{eff}(\omega) \cdot \omega = \begin{cases}
\gamma_1\omega & -\omega_{db} \leq \omega \leq \omega_{db} \\
\gamma_2\omega + (\gamma_2 - \gamma_1)\omega_{db} & \omega \leq -\omega_{db} \\
\gamma_2\omega + (\gamma_1 - \gamma_2)\omega_{db} & \omega \geq \omega_{db} \\
\end{cases}
    \label{eq:gamma_model}
\end{align}
where $\omega_{db}$ is the deadband frequency range and $\gamma_1$ is the damping term within the deadband and $\gamma_2$ outside of deadband.

\begin{figure}[ht!]
  \centering
  \includegraphics[width=\columnwidth]{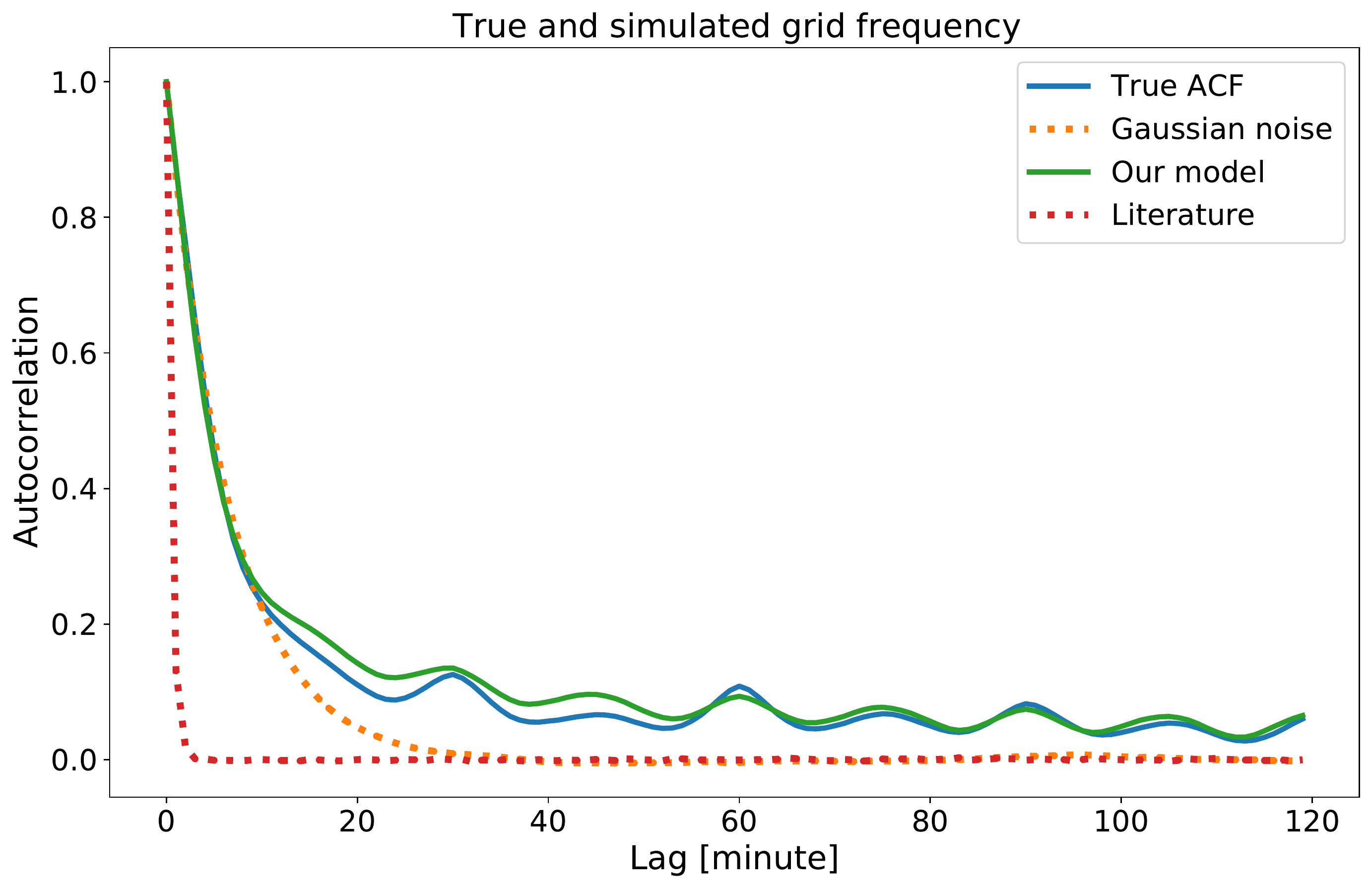}
  \caption{True ACF compared with a simple OU model with Gaussian noise (model class as in \cite{stochastic_tpwrs, g2, g3, oup_load, double_peaked_1}), and our model based on (\ref{eq:freq_full}) and Table\,\ref{tab:table2}. A specific model from literature \cite{double_peaked_2} is selected here to emphasise that while it matches the PDF well (Fig.\ref{fig:hist_simulation}) it does not capture the time dynamics of grid frequency (encoded in ACF) at all.}
  \label{fig:acf_simulations}
\end{figure}

\begin{figure}[ht!]
  \centering
  \includegraphics[width=\columnwidth]{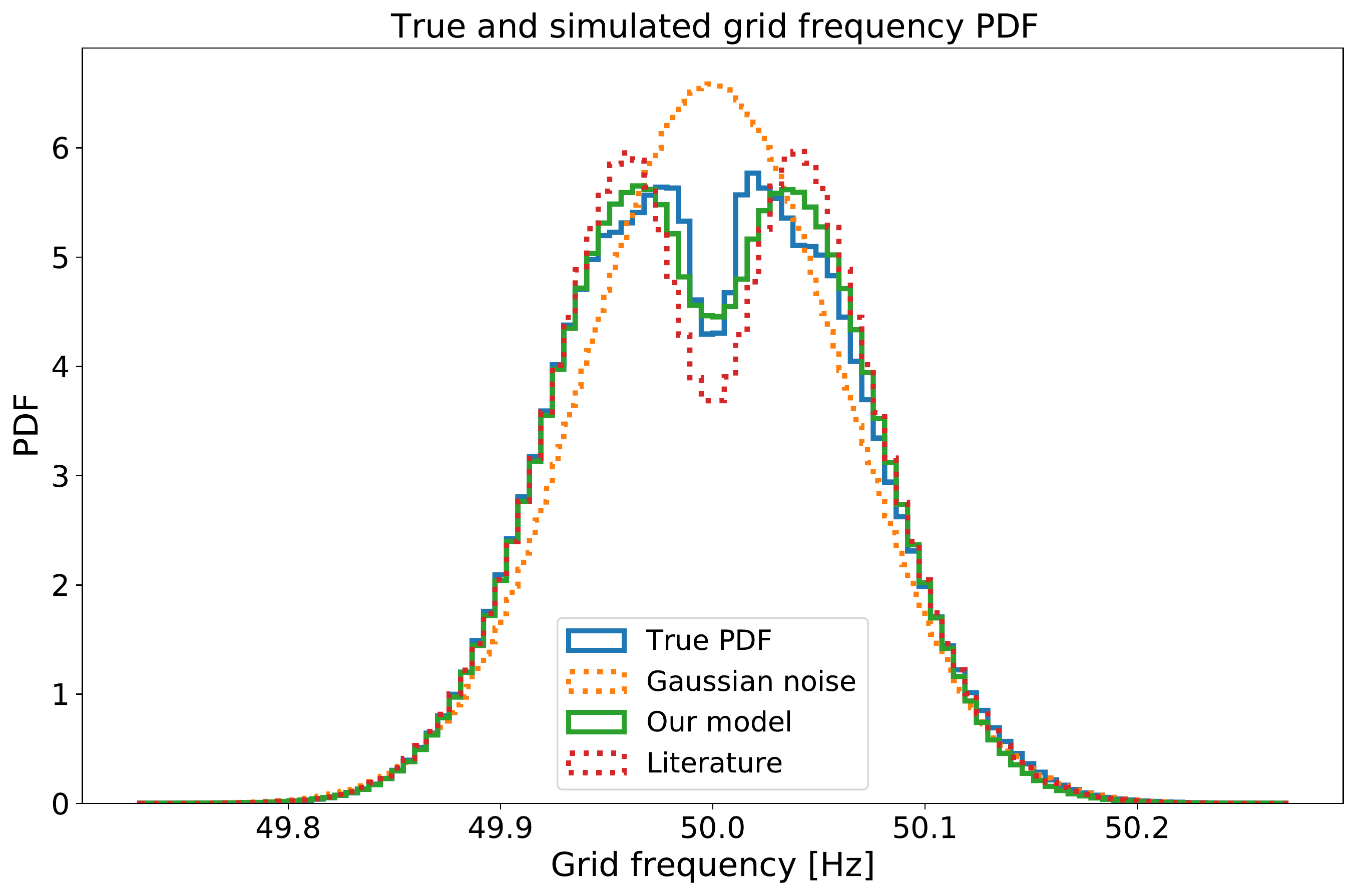}
  \caption{True PDF compared with a simple OU model with Gaussian noise, and a model based on (\ref{eq:freq_full}) and Table\,\ref{tab:table2}, and a model from literature \cite{double_peaked_2}. All models match the standard deviation and the mean of the actual frequency data.}
  \label{fig:hist_simulation}
\end{figure}

\begin{figure}[ht!]
  \centering
  \includegraphics[width=\columnwidth]{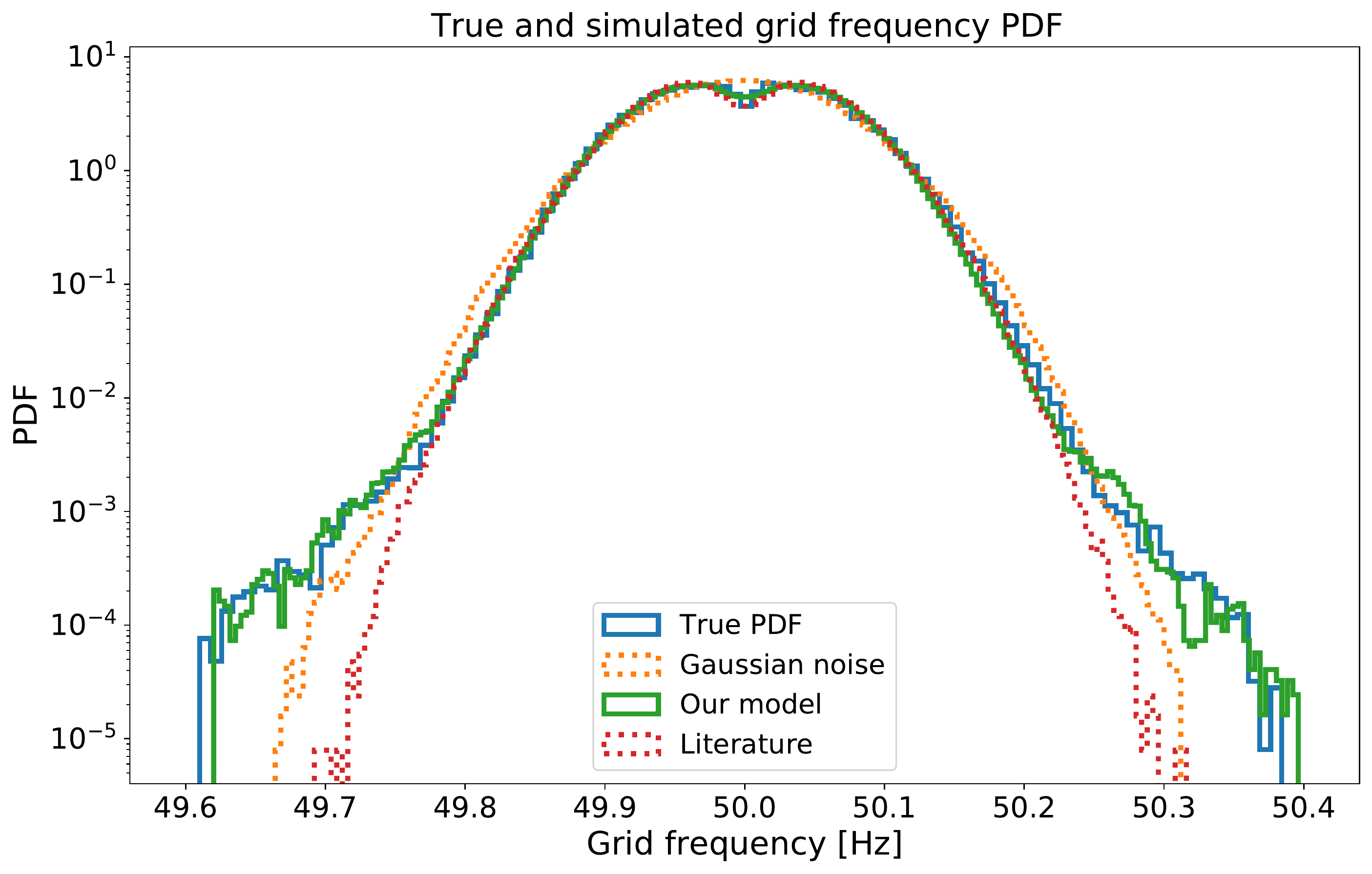}
  \caption{Log plot of Fig.\,\ref{fig:hist_simulation} emphasising the tail behaviour.}
  \label{fig:hist_simulation_tails}
\end{figure}

\begin{table}[!t]
\caption{Model parameters \label{tab:table2}}
\centering
\begin{tabular}{|c||c|}
\hline
Parameter & Value \\
\hline
\hline
Hurst exponent H & 0.75 \\
\hline
L\'{e}vy $\alpha$ & 1.975 \\
\hline
$\omega_{db}$ & 15\, mHz\\
\hline
$\gamma_1$ & 0 \\
\hline
$\gamma_2$ & 1/150 \\
\hline
$\sigma$ & 0.0021 \\
\hline
$\mu(t)$ & as in Fig.\ref{fig:rocof_fold} \\
\hline
\end{tabular}
\end{table}
With our model we can reproduce both the exponential drop-off and power-law tail in the ACF, as well as periodic structure (see Fig.\,\ref{fig:acf_simulations}). We can also reproduce the over-all scale of fluctuations and the double-peaked structure (Fig.\,\ref{fig:hist_simulation}). Moreover, our model also reproduces the heavy tails of the PDF (Fig.\,\ref{fig:hist_simulation_tails}). This was achieved with the parameter combination in Table\,\ref{tab:table2}. 

Besides our model we reproduced the results of \cite{double_peaked_2}, labelled `Literature'. This particular model was selected to illustrate the importance of looking at all statistical properties together. `Literature' model agrees well with the PDF (Fig.\,\ref{fig:hist_simulation}) but does not capture the timeseries properties of the grid frequency at all (Fig.\,\ref{fig:acf_simulations}). For the class of simple baseline OU models with non-fractional Gaussian noise, labelled `Gaussian noise' in the figures, we chose constant $\gamma^{eff}(\omega)=1/400$ such that the initial exponential drop in the ACF is correctly captured and the overall scale $\sigma=0.064$ of fluctuations is reproduced. Note that all models based on the swing equation and non-fractional Gaussian noise will fail to reproduce long-term correlations as seen in the ACF and the power spectrum even if the overall PDF and short term ACF fit well.

\subsection{Applications}
{More elaborate modelling of the grid frequency, compared with a simple OUP (i.e. non-fractional Gaussian noise), should lead to improvements in real-life applications of the models. 

Frequency response services are automatic balancing services provided by devices (batteries, diesel generators, industrial HVACs,...) to an ESO which are controlled by the value of the grid frequency. In general, in GB and elsewhere, there are two types of frequency response services: \emph{dynamic}, where power generation/consumption of an asset is proportional to the deviation in the grid frequency after some trigger, and \emph{static}, where power is generated/consumed at a fixed value after the service is triggered at some frequency deviation. If the devices delivering these services do not deliver the required energy for the required duration then strict financial penalties usually apply\footnote{See \url{https://www.nationalgrideso.com/industry-information/balancing-services/frequency-response-services/firm-frequency-response-ffr}}. 
\paragraph{Dynamic frequency response}
Devices participating in the dynamic service (such as batteries) need to know the yearly distribution of energy requirements. The distribution is directly related to the timeseries of frequency fluctuations and is used to evaluate the ability of the device to deliver the service (or calculate the percentage of times the device will fail to deliver) as well as calculate the compensation for the provision of service as it is directly linked to the energy delivered.

We take the parameters for a typical GB high-frequency dynamic response service, where the proportional response is required within a couple of seconds for deviations above 50.015\,Hz.
\begin{figure}[ht!]
  \centering
  \includegraphics[width=\columnwidth]{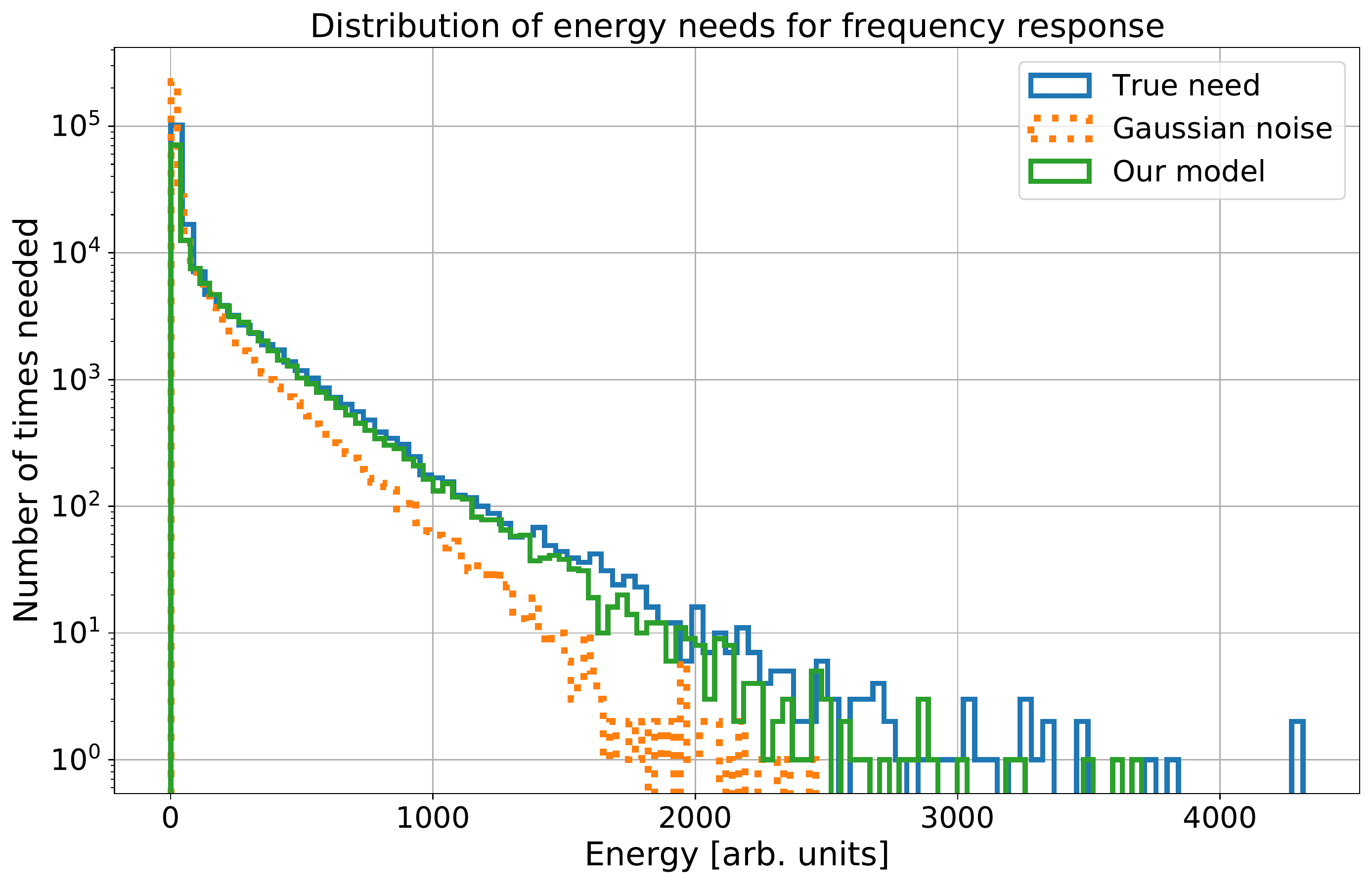}
  \caption{Distribution of energy requirement in a year for a dynamic high frequency response as determined by our models vs true requirement and a model based on Gaussian noise.}
  \label{fig:ffr_requirement_e}
\end{figure}
 In Fig.\,\ref{fig:ffr_requirement_e} we plot the distribution of energy requirements for one year. We see that the Gaussian OUP, despite matching the over scale of fluctuations (see Figs.\,\ref{fig:hist_simulation}, \ref{fig:acf_simulations}), underestimates larger energy requirements, due to the lack of long-term correlations in deviations from the nominal value, whereas our model reproduces the true requirement. This could lead to the wrong sizing of the battery destined for this service, wrong estimation of expected income, and higher than expected failure rate to delivery required energy. For example, looking at Fig.\,\ref{fig:ffr_requirement_e}, using a simple OUP model to determine the size of a battery (determined by the largest energy requirement in a year) the owner of the device would size the battery to have capacity at 2000 (arb. units) but true requirement is closer to 3000 (arb. units).

\paragraph{Static frequency response}
Similarly as above, evaluation of a device participating in a static delivery of frequency response requires the knowledge of the distribution of response durations after each trigger. We take
parameters for a typical low-frequency static response service, where a fixed power is generated within a couple of seconds for deviations below the  49.9\,Hz trigger. 
\begin{figure}[ht!]
  \centering
  \includegraphics[width=\columnwidth]{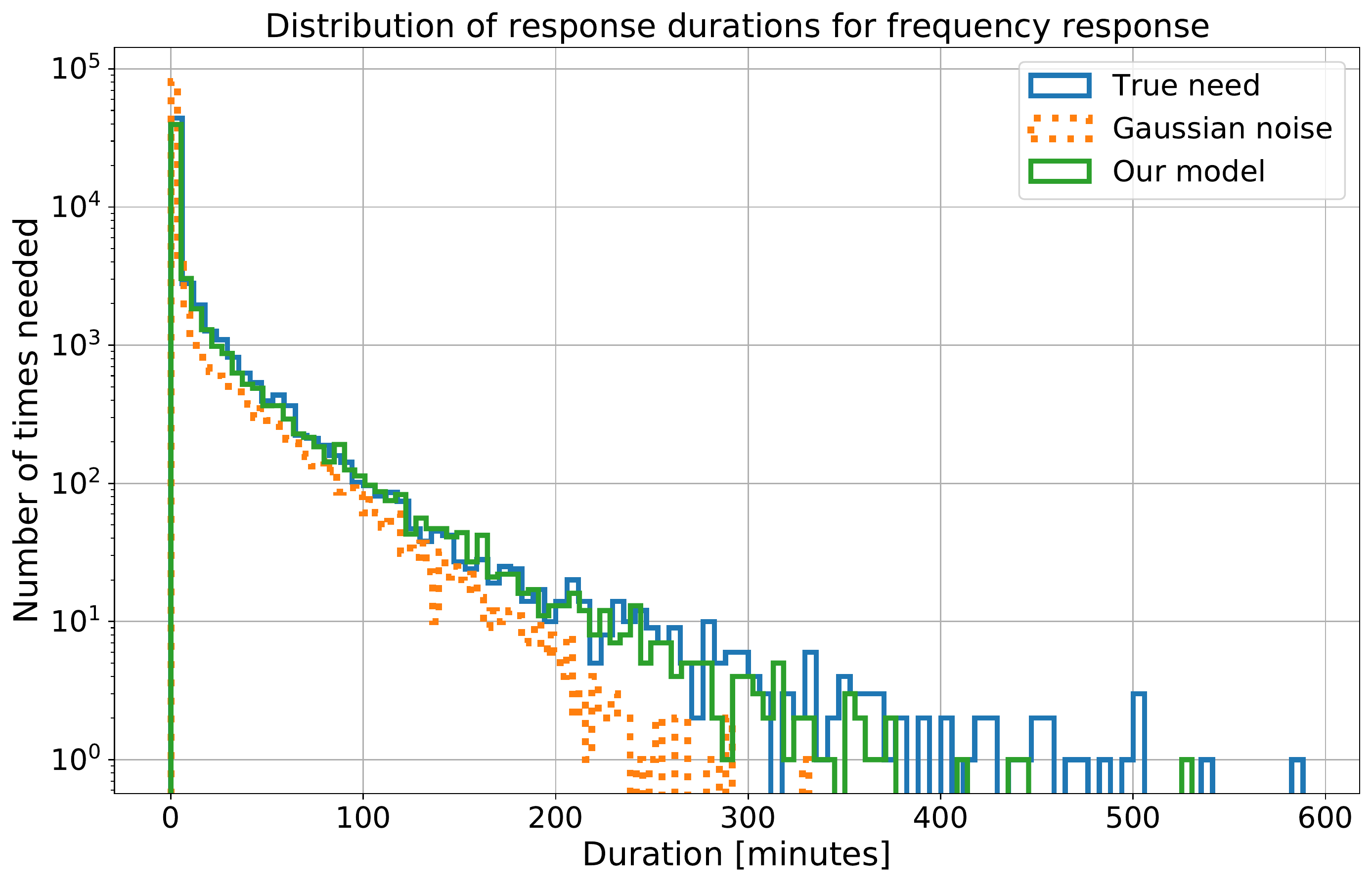}
  \caption{Distribution of response durations in a year for a static low frequency response as determined by our models vs true requirement and a model based on Gaussian noise.}
  \label{fig:ffr_requirement_d}
\end{figure}
In Fig.\,\ref{fig:ffr_requirement_d} we plot the required response durations and again see that Gaussian OUP underestimates the frequency of the longer responses. For example, diesel generators that are committed to provide static response need to provide responses for the required length but are also limited to how long they can run in each year\footnote{See e.g. \url{https://ec.europa.eu/environment/industry/stationary/mcp.htm}} for environmental reasons. Using a simple OUP would underestimate the frequency of response durations (and consequently the amount of fuel needed) by more than a factor of two for events that would need responses longer that 2 hours (see Fig.\,\ref{fig:ffr_requirement_d}).

\paragraph{Static frequency response and inertia change}
Swing equation enables us, for example, the evaluation of what would requirement be for e.g. static frequency response if inertia in the grid changed with all other things being equal. Looking back at the derivation of the swing equation (\ref{eq:freq_phys}) and at the form of the equation with explicit inertia constant (\ref{eq:inert_droop}) we see that lower inertia (while everything else stays the same) can be modelled by reducing the damping factor $\gamma$ \emph{and} size of fluctuations in $\pi(t)$ by the same factor. In Fig.\,\ref{fig:ffr_requirement_i} we compare two response duration distributions predicted by our model with differing inertia (by 20\%).
\begin{figure}[ht!]
  \centering
  \includegraphics[width=\columnwidth]{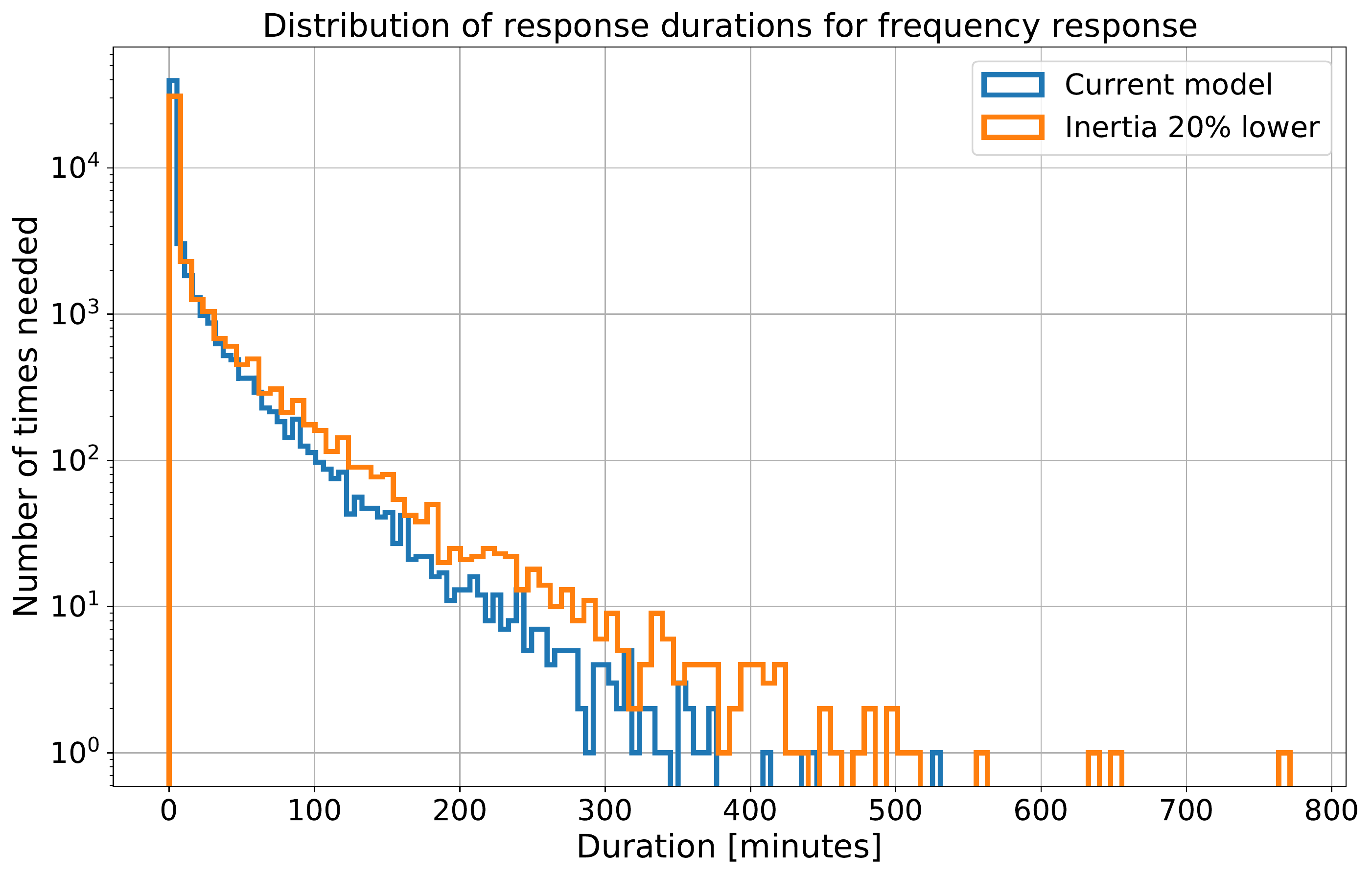}
  \caption{Distribution of response durations in a year for a static low frequency response as determined by our model for differing grid inertia. Note that if the size of intrinsic fluctuations stays the same while the inertia decreases, the overall fluctuations in the grid frequency will increase (\ref{eq:inert_droop}). Because of long term correlations large deviations will persist for longer, requiring more frequent longer responses to frequency deviations.}
  \label{fig:ffr_requirement_i}
\end{figure}
We see that lower inertia would mean more frequent long responses, as frequency deviations would stay longer below trigger level. For example, responses long about 2 hours would be twice as likely if overall yearly inertia would drop by 20\% (all other things being equal).}

\section{Conclusions}
We study the statistical properties of the grid frequency, focusing on the data for the power system of Great Britain. We show that some curious statistical properties such as the double-peaked probability density function, heavy tails, periodicity, and the long-range power-law dependence in the autocorrelation function, can be modelled successfully by slight modifications of the swing equation or equivalently the Ornstein-Uhlenbeck stochastic process. We empirically show that the effective damping factor is frequency-dependent, which reproduces the double-peakedness. We show that a better description for the underlying noise process, compared to Gaussian, is fractional noise, which due to inherent long term correlations between noise increments reproduces the power-law behaviour in the autocorrelation function. Additionally, we source the fractional noise from heavy-tailed distribution to correctly capture the fat tails of the grid frequency probability density function. Finally, we look at applications of our model. We determine the distribution of energy requirement for a dynamic frequency response service, response duration distribution for a static frequency response service and show their predictions are closer to true values compared to predictions from a standard swing equation model.

\section{Acknowledgments}
\noindent The author would like to thank the reviewers. Their suggestions have vastly improved the initial version of this contribution.

%\clearpage
\appendix[Statistical properties of other grids]
\subsection{European Continental Synchronous grid}
\begin{figure}[ht!]
  \centering
  \includegraphics[width=\columnwidth]{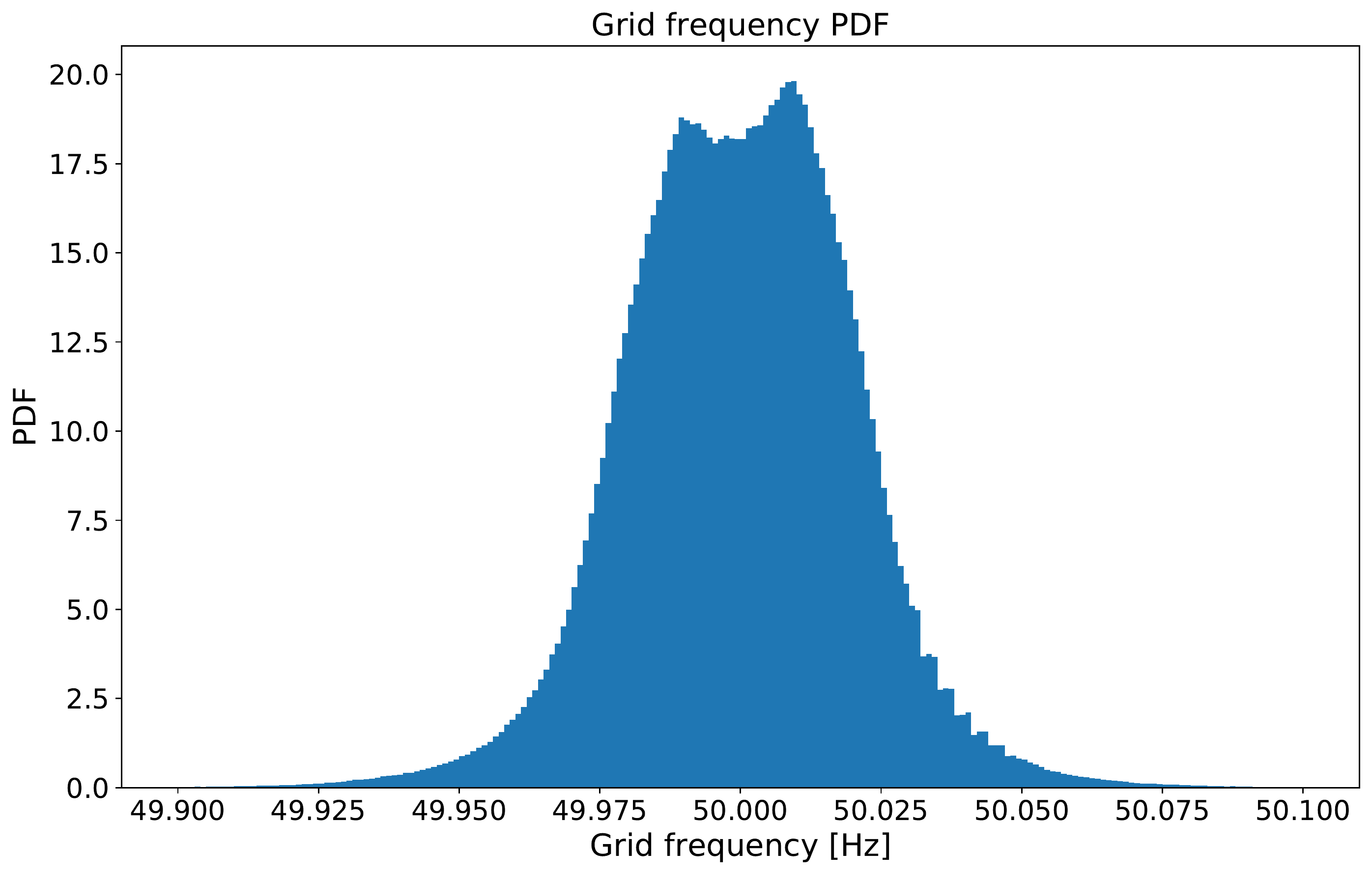}
  \caption{European grid frequency PDF. Note the slight double-peaked nature.}
  \label{fig:hist_CE}
\end{figure}

\begin{figure}[ht!]
  \centering
  \includegraphics[width=\columnwidth]{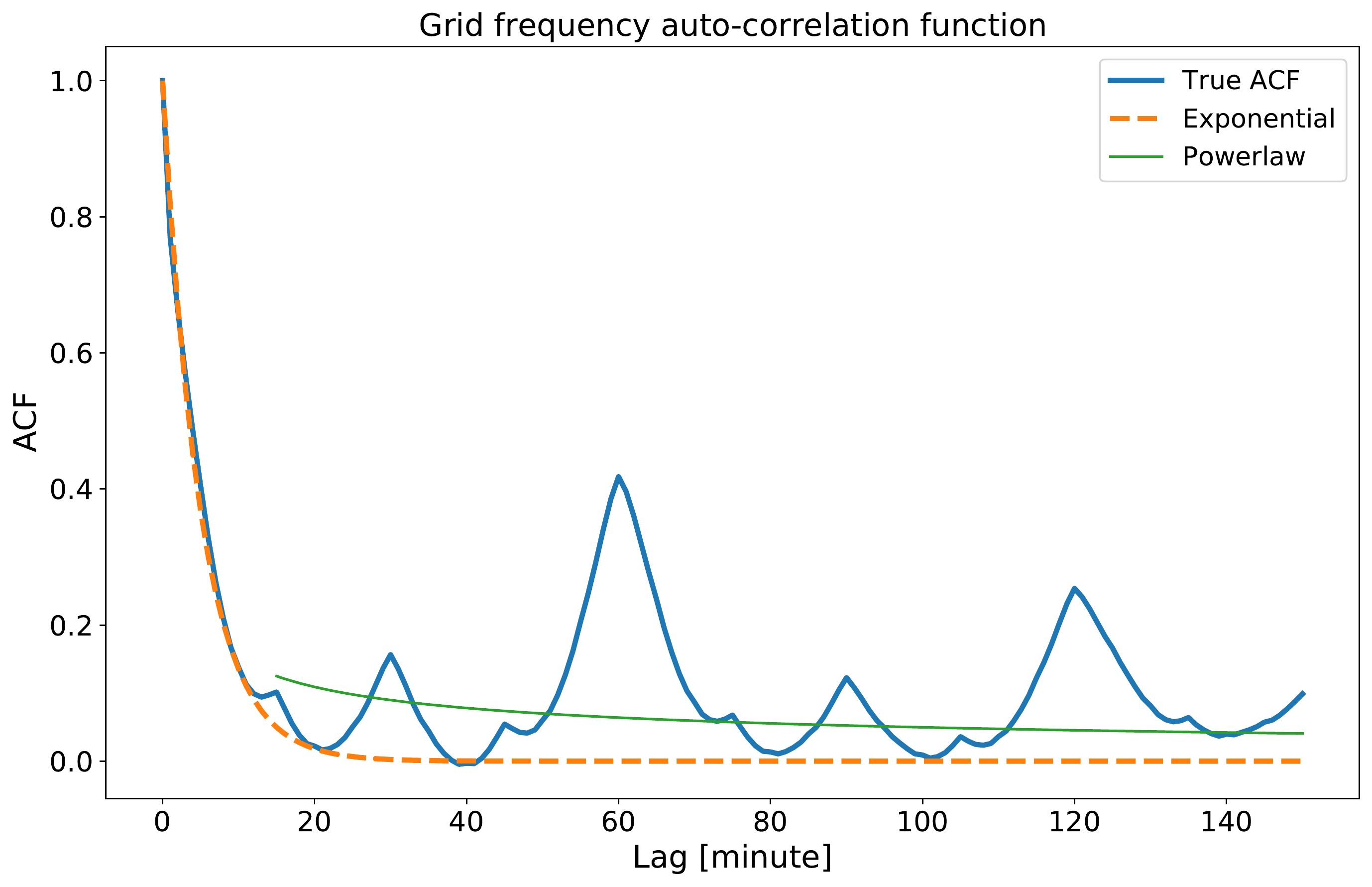}
  \caption{European grid frequency ACF. Note the strong periodicity and power-law behaviour at longer times.}
  \label{fig:ACF_CE}
\end{figure}

\begin{figure}[ht!]
  \centering
  \includegraphics[width=\columnwidth]{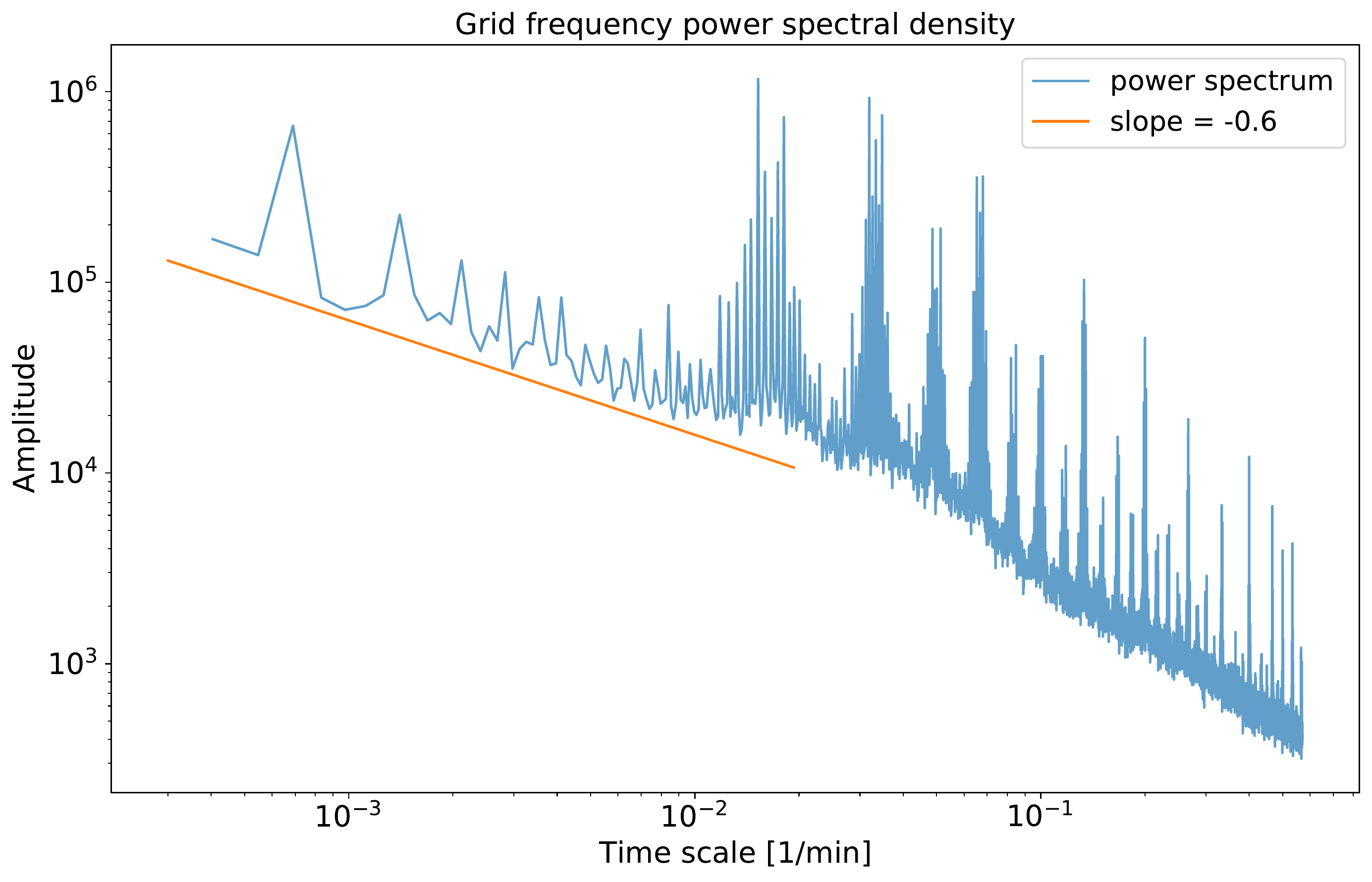}
  \caption{European grid power spectrum. Note the non-zero slope at long time scales signalling long-term correlated noise (as opposed to simple Gaussian).}
  \label{fig:PS_CE}
\end{figure}

{We show that the European grid exhibits similar properties to the GB frequency: double-peakedness (Fig.\,\ref{fig:hist_CE}), periodicity and long term correlations suggesting fractional noise fluctuations (Figs.\,\ref{fig:ACF_CE}, \ref{fig:PS_CE}). Therefore same modelling approaches as in this paper can be used for the European grid. The data was obtained from the French ESO\cite{rte_data} and consists of measurements sampled every 10 seconds. We study the data from 2017 to 2020 inclusive.}

\subsection{Nordic Grid}
\begin{figure}[ht!]
  \centering
  \includegraphics[width=\columnwidth]{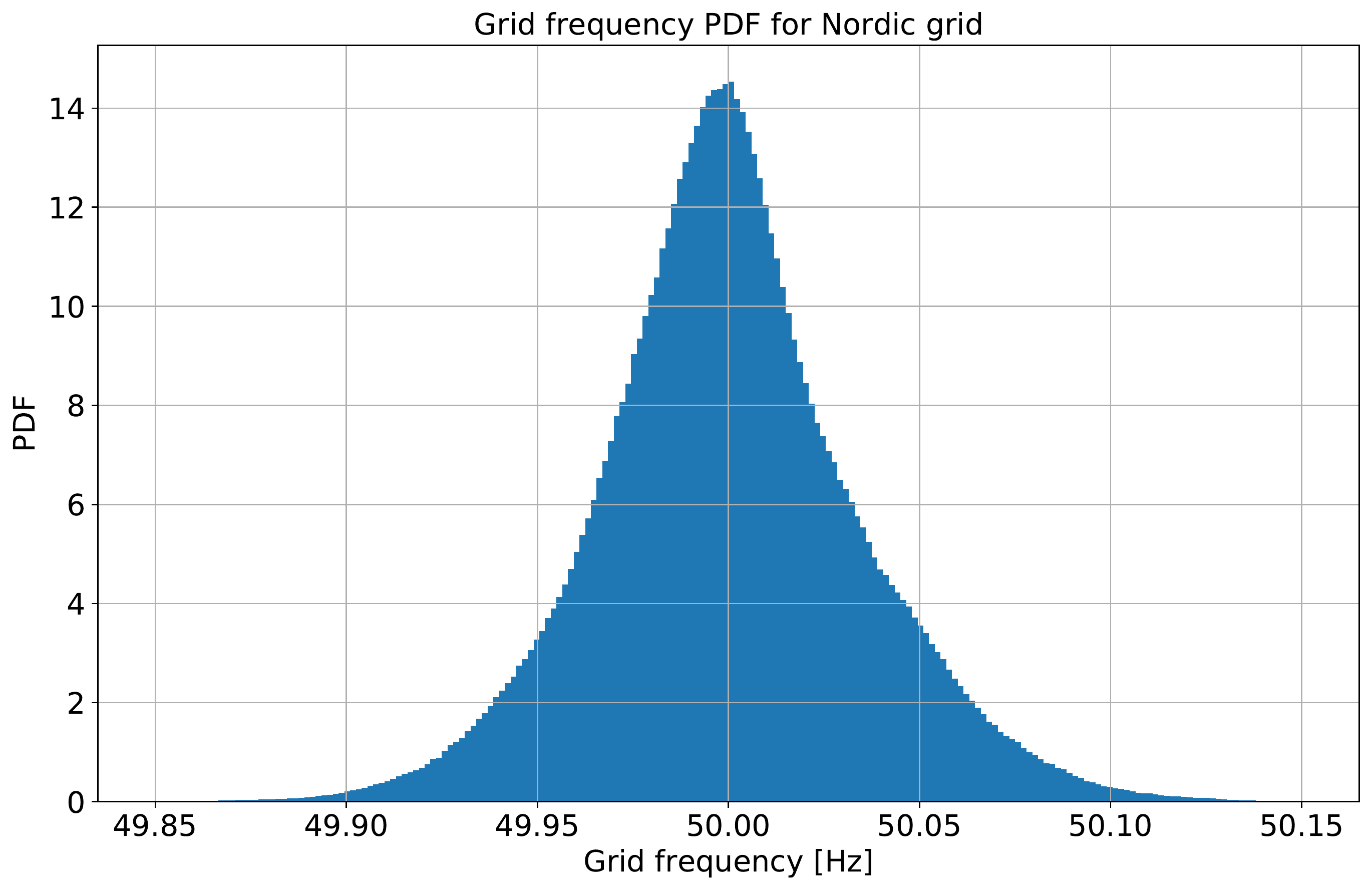}
  \caption{Nordic grid frequency PDF. Note the uni-modal distribution.}
  \label{fig:hist_NORDIC}
\end{figure}

\begin{figure}[ht!]
  \centering
  \includegraphics[width=\columnwidth]{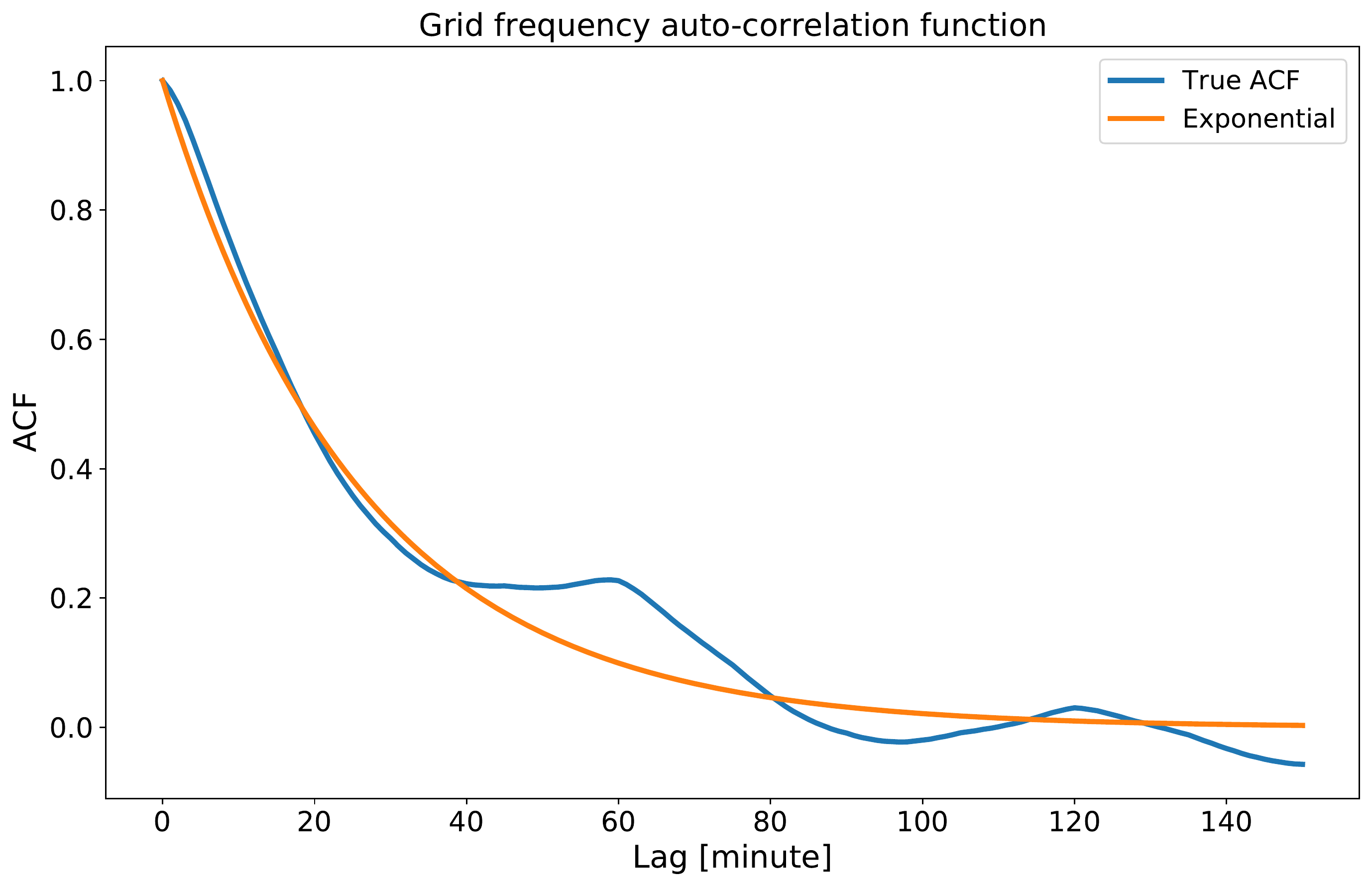}
  \caption{Nordic grid frequency ACF. Note that despite strong correlations at about 40min the ACF is well-described by an exponential, signalling swing equation + Gaussian noise is a good description.}
  \label{fig:ACF_NORDIC}
\end{figure}

\begin{figure}[ht!]
  \centering
  \includegraphics[width=\columnwidth]{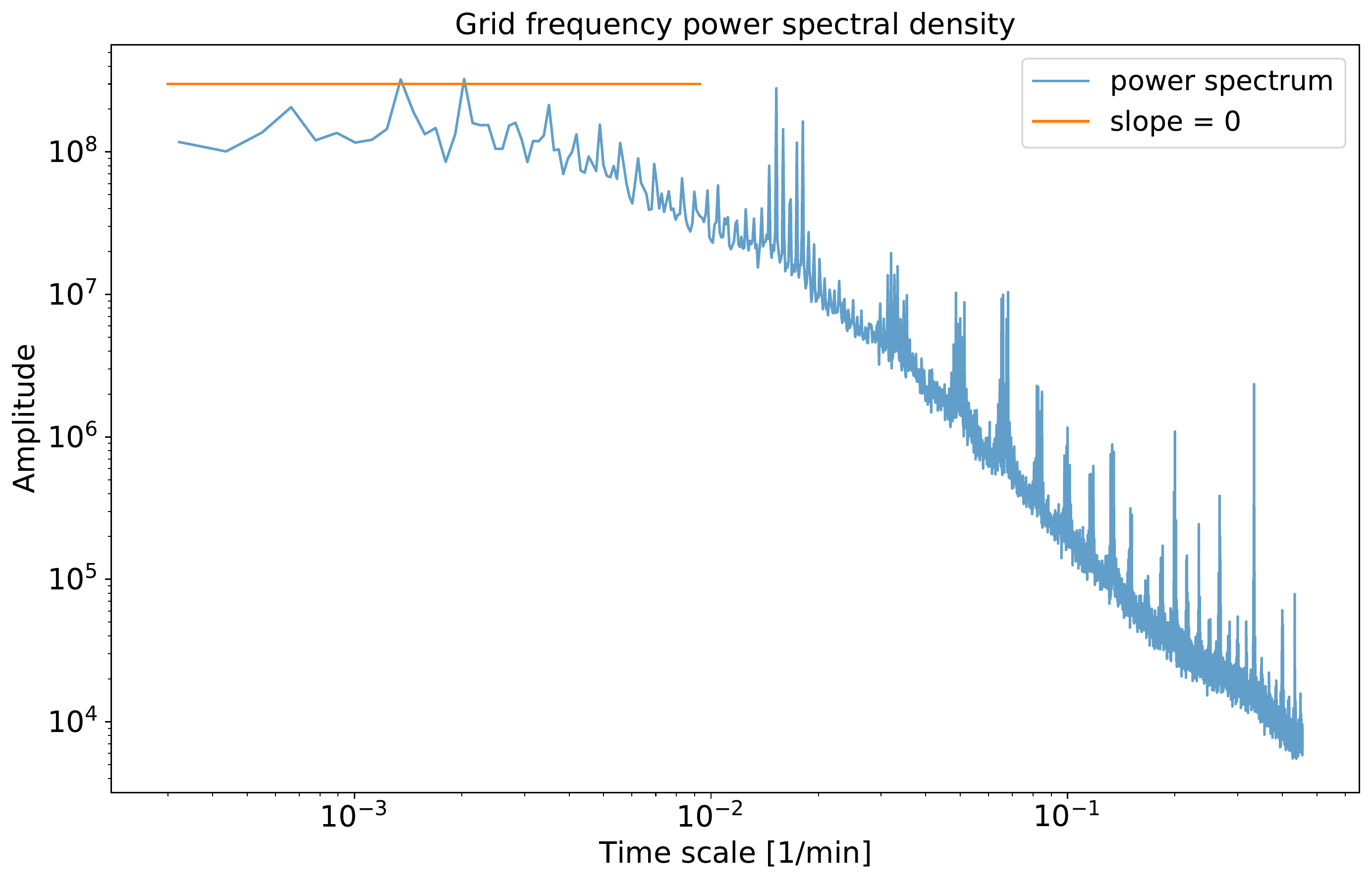}
  \caption{Nordic grid frequency power spectrum. Note that at large time-scales the power spectrum is flat as expected for OUP (see (\ref{eq:psd})).}
  \label{fig:PS_NORDIC}
\end{figure}

{We show in Fig.\,\ref{fig:hist_NORDIC} the PDF for the Nordic grid is uni-modal and that despite strong correlations at longer times the ACF (Fig.\,\ref{fig:ACF_NORDIC}) is well described by an exponential, suggesting ordinary Gaussian fluctuation. This is also confirmed by checking the large time-scale part of the frequency power spectrum (Fig.\,\ref{fig:PS_NORDIC}). The data for the Nordic synchronous area is obtained from the Norwegian ESO\cite{statnett_data} and consists of second by second measurements of the frequency for the year 2020.}

\subsection{Mallorcan Grid}
\begin{figure}[ht!]
  \centering
  \includegraphics[width=\columnwidth]{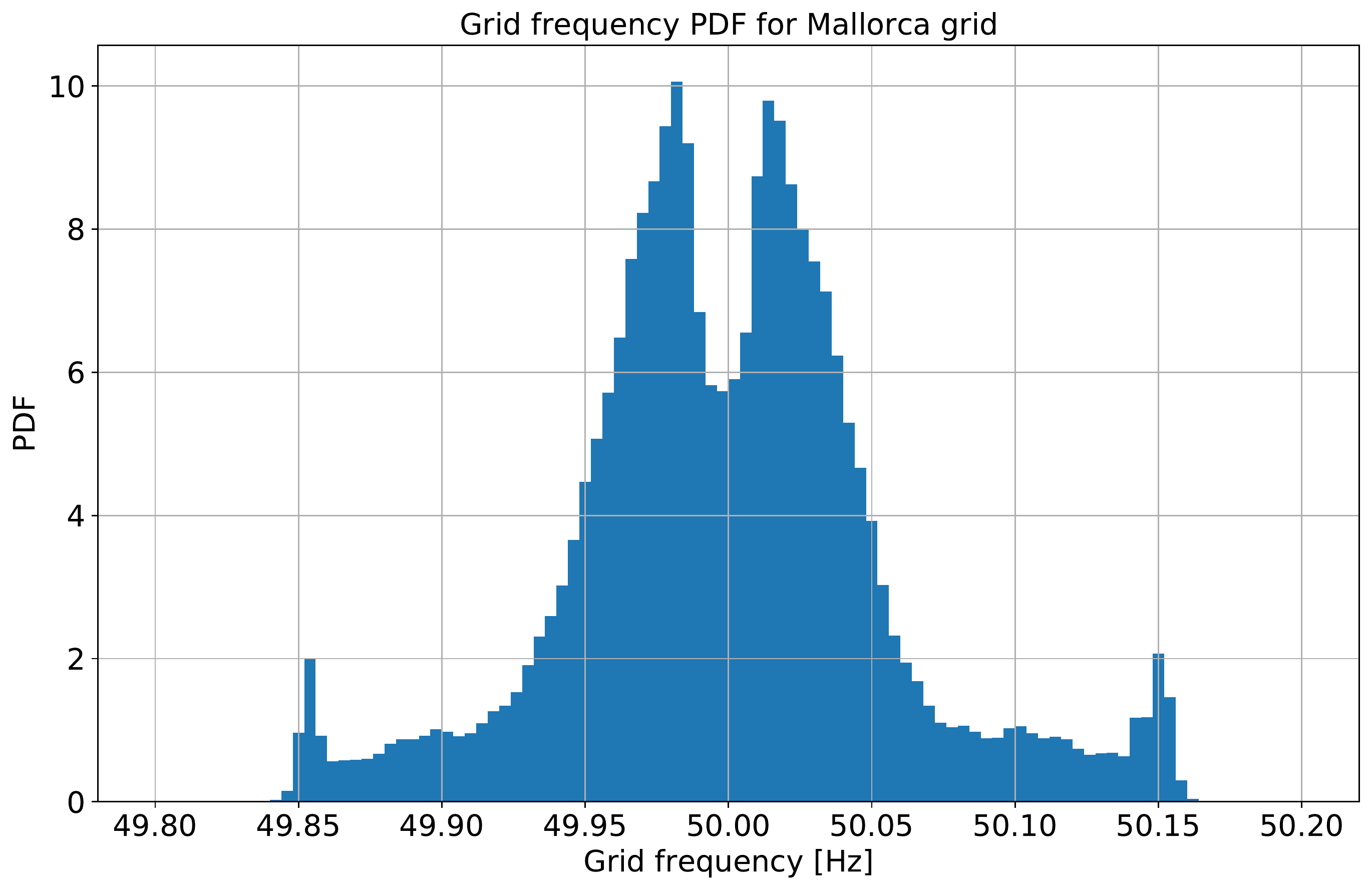}
  \caption{Mallorcan grid frequency PDF. Note the multimodal structure suggesting additional features in the restoring term of (\ref{eq:freq_full}), in addition to the deadband structure.}
  \label{fig:hist_MAL}
\end{figure}

\begin{figure}[ht!]
  \centering
  \includegraphics[width=\columnwidth]{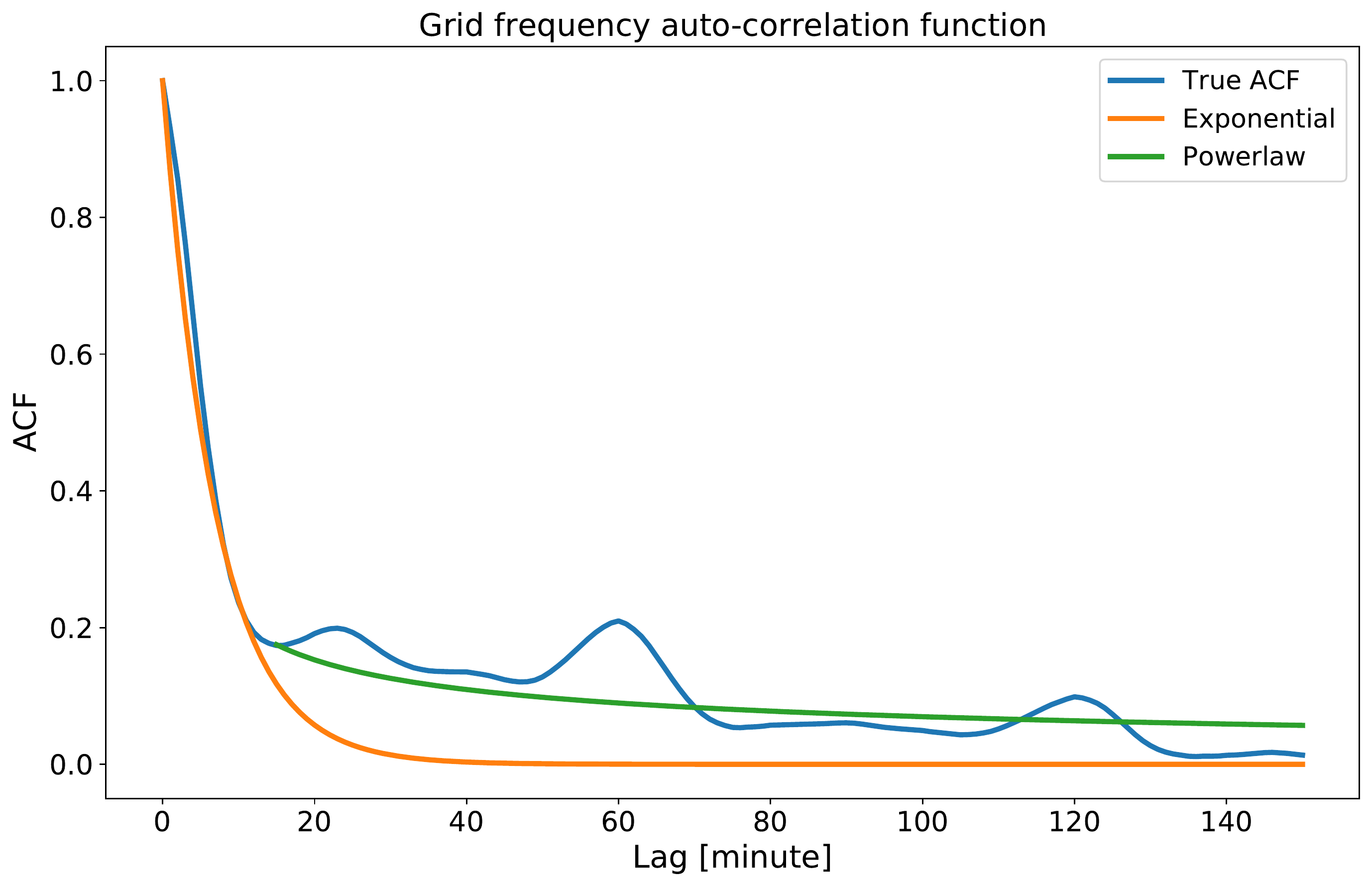}
  \caption{Mallorcan grid frequency ACF. Note how the initial fall in correlations can be described by an exponential, but at later times it is better described by a power-law, suggesting fractional noise.}
  \label{fig:ACF_MAL}
\end{figure}

\begin{figure}[ht!]
  \centering
  \includegraphics[width=\columnwidth]{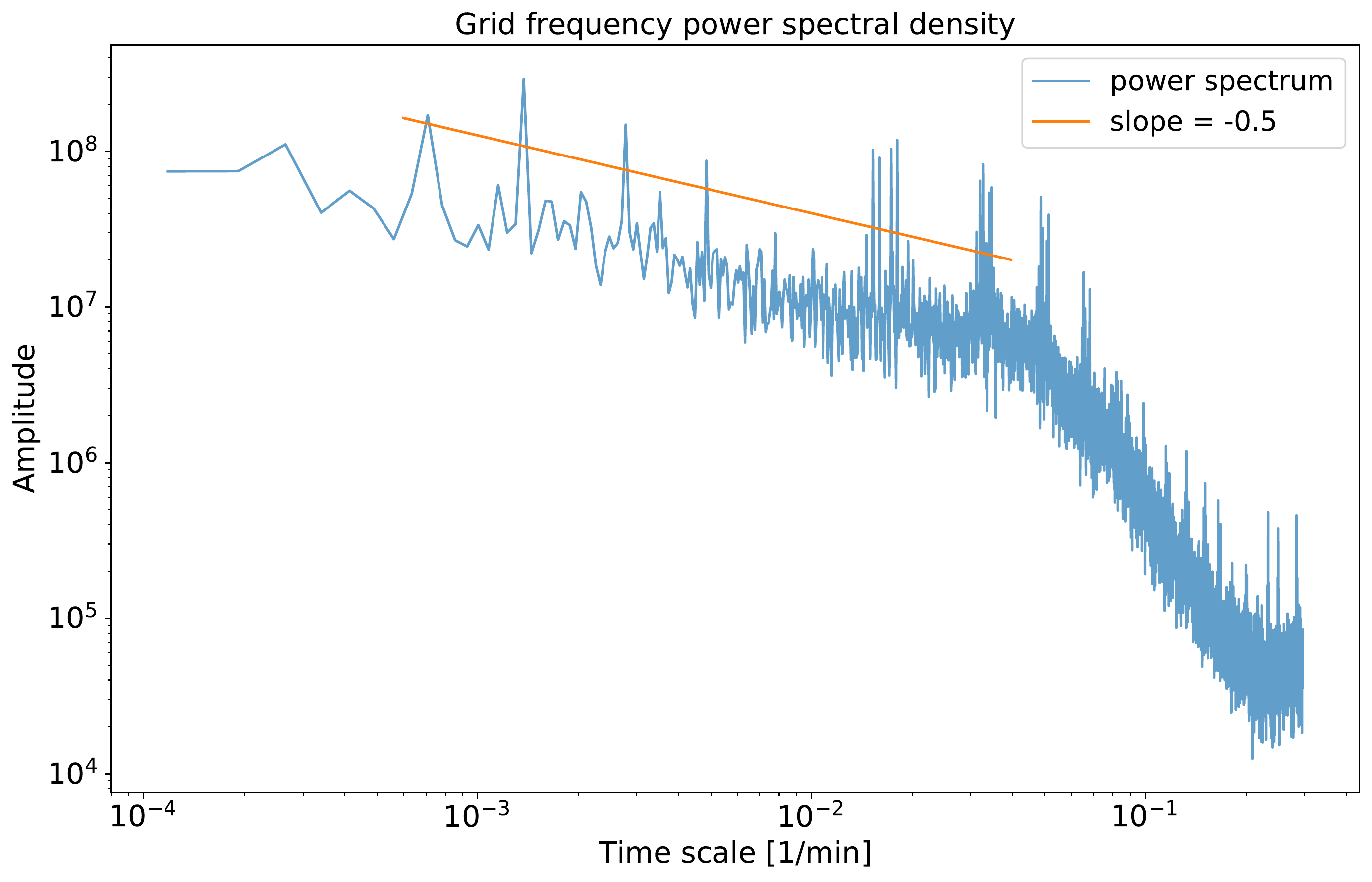}
  \caption{Mallorcan grid frequency power spectrum. Note the non-trivial slope at large time scales suggesting fractional noise is a better description than Gaussian.}
  \label{fig:PS_MAL}
\end{figure}

{Mallorcan grid is a small island grid that in addition to the two peaks around the central value (as in GB, Europe) exhibits extra structure in frequency PDF (Fig.\,\ref{fig:hist_MAL}). Similarly as for GB and European grid we show that there are long term power-law correlations suggesting fractional noise fluctuations (Figs.\,\ref{fig:ACF_MAL}, \ref{fig:PS_MAL}). The data was obtained from \cite{database} and consists of 94 days of seconds-by-second measurements in 2019.}

% Can use something like this to put references on a page
% by themselves when using endfloat and the captionsoff option.
\ifCLASSOPTIONcaptionsoff
  \newpage
\fi

% trigger a \newpage just before the given reference
% number - used to balance the columns on the last page
% adjust value as needed - may need to be readjusted if
% the document is modified later
%\IEEEtriggeratref{8}
% The "triggered" command can be changed if desired:
%\IEEEtriggercmd{\enlargethispage{-5in}}

% references section

% can use a bibliography generated by BibTeX as a .bbl file
% BibTeX documentation can be easily obtained at:
% http://mirror.ctan.org/biblio/bibtex/contrib/doc/
% The IEEEtran BibTeX style support page is at:
% http://www.michaelshell.org/tex/ieeetran/bibtex/
\bibliographystyle{IEEEtran}
% argument is your BibTeX string definitions and bibliography database(s)
\bibliography{IEEEabrv,bare_jrnl}
%
% <OR> manually copy in the resultant .bbl file
% set second argument of \begin to the number of references
% (used to reserve space for the reference number labels box)
% \begin{thebibliography}{1}

% \bibitem{IEEEhowto:kopka}
% H.~Kopka and P.~W. Daly, \emph{A Guide to \LaTeX}, 3rd~ed.\hskip 1em plus
%   0.5em minus 0.4em\relax Harlow, England: Addison-Wesley, 1999.

% \end{thebibliography}

% biography section
% 
% If you have an EPS/PDF photo (graphicx package needed) extra braces are
% needed around the contents of the optional argument to biography to prevent
% the LaTeX parser from getting confused when it sees the complicated
% \includegraphics command within an optional argument. (You could create
% your own custom macro containing the \includegraphics command to make things
% simpler here.)
%\begin{IEEEbiography}[{\includegraphics[width=1in,height=1.25in,clip,keepaspectratio]{mshell}}]{Michael Shell}
% or if you just want to reserve a space for a photo:

% \begin{IEEEbiography}{Michael Shell}
% Biography text here.
% \end{IEEEbiography}

% % if you will not have a photo at all:
\begin{IEEEbiographynophoto}{David Kraljic} 
received an MSci in experimental and theoretical physics from the University of Cambridge in 2012 and a DPhil in theoretical physics from the University of Oxford in 2016.

He works as a researcher in the Laboratory of Robotics, Faculty of Electrical Engineering, University of Ljubljana from 2017 focusing on gait analysis as well as industrial robotics. He also works at Comcom trading d.o.o., an electricity trading and technology company, where he focuses on power system modelling and optimisation.
\end{IEEEbiographynophoto}

% insert where needed to balance the two columns on the last page with
% biographies
%\newpage

% \begin{IEEEbiographynophoto}{Jane Doe}
% Biography text here.
% \end{IEEEbiographynophoto}

% You can push biographies down or up by placing
% a \vfill before or after them. The appropriate
% use of \vfill depends on what kind of text is
% on the last page and whether or not the columns
% are being equalized.

%\vfill

% Can be used to pull up biographies so that the bottom of the last one
% is flush with the other column.
%\enlargethispage{-5in}

% that's all folks
\end{document}